\documentclass[aps,prb,showpacs,twocolumn,groupedaddress]{revtex4-1}
\usepackage{amsmath}
\usepackage{cancel}
\usepackage{float}
\usepackage{ifsym}
\usepackage{color}
\usepackage[dvipsnames]{xcolor}
\usepackage{latexsym}
\usepackage{latexsym}
\usepackage{dcolumn}
\usepackage{epsfig}
\usepackage{amsfonts}
\usepackage{amsmath}
\usepackage{dsfont}
\usepackage{accents}
\usepackage[all,cmtip]{xy}
\usepackage{paralist}
\usepackage{enumerate}
\usepackage{amssymb}
\usepackage{bm}
\usepackage{graphicx}
\usepackage{wasysym}
\usepackage{hyperref}

\begin{document}

\title{Polarity of domain boundaries in nonpolar materials derived from order parameter and layer group symmetry}
\author{W.~Schranz$^1$}
\email{wilfried.schranz@univie.ac.at}
\affiliation{$^1$University of Vienna, Faculty of Physics, Boltzmanngasse 5, A-1090 Wien, Austria} 
\author{I.~Rychetsky$^2$} \author{J.~Hlinka$^2$} 
\affiliation{$^2$Institute of Physics, Academy of Sciences of the Czech Republic, Na Slovance 2, 18221 Prague 8, Czech Republic.}

\date{\today}

\begin{abstract}
Domain boundaries and other twin boundaries in crystalline materials are receiving increasing interest. They can carry unique functional properties, which in many cases are absent in the surrounding bulk material. One such property of domain boundaries can be their electric polarity. Phenomenological insight in the polarity of domain boundaries was so far based either on the knowledge of the order parameter and the form of Landau-Ginzburg free energy functional, or on the knowledge of the symmetry of the domain boundaries. 
In the present work we show on the concrete examples of potassium thiocyanate (KSCN) and lacunar spinel crystals, that the concept of the primary order-parameter can help to find the layer group describing the maximal possible symmetry of a given domain boundary. Combination of layer group and order parameter symmetries is then employed to clarify the nature of the polarity of domain boundaries.
\end{abstract}


\maketitle

\section{Introduction} \label{sec:Introduction}

Domains in ferroelectric and ferromagnetic crystals are well known for their applications in microelectronic devices \cite{Bain2017,Spaldin2012,Wadhawan2000}. Domains are three dimensional objects (with 3d translational symmetry), which usually appear due to breaking of crystal symmetry at a structural phase transition. The experimental investigation and theoretical description of domains and their properties has a long and fruitful tradition \cite{Tagantsev2010}. On the other hand, domain  boundaries  or boundaries in general have been recognized as useful objects much later. Planar domain walls are objects with 2d - translational periodicity, which separate adjacent domain states, homogeneous in 3d.  
Thanks to the enormous progress in the development of high resolution techniques, local structures of domain boundaries are nowadays explored in very detail \cite{Seidel2009,Seidel2016}. In addition, also local properties of domain boundaries are measured. This has led to fascinating discoveries, e.g. of superconducting twin boundaries in  WO$_{3}$ \cite{Aird1998}, conducting domain walls in insulating BaTiO$_{3}$ \cite{Sluka2013} or polarity of domain boundaries of non-polar perovskites CaTiO$_{3}$ \cite{Aert2012,Yokota2014,Yokota2017}, SrTiO$_{3}$ \cite{Salje2013}, LaAlO$_{3}$ \cite{Salje2016} and PbZrO$_{3}$ \cite{Wei2014}.

In this paper, a  twin boundary is considered  to be the interface between two equivalent structural variants of a chemically homogeneous crystalline material, such that these variants can be superposed by a combination of euclidean translations and proper or improper rotations.  We use domain wall or boundary as a special case of the twin boundary, separating the structural variants (transformation twins) related by some symmetry operations of a parent high-symmetry phase. For our purposes the interface is mostly understood as a flat object with a negligible curvature, and although there is no sense to define a complete, total  thickness of the domain boundary, we can have in mind a layer of the structure with a measurably different structure than the adjacent bulk domains, typically of the order of the domain wall thickness estimated from Landau-Ginzburg models.

Various theoretical approaches for the description of domain boundaries exist. They are based on pure geometrical arguments using layer groups \cite{Janovec1976,Janovec1981,Janovec2006,Janovec1997,Janovec1989,Kopsky2008,Janovec2004,Janovec2011,Privratska1997,Privratska2000}, 
Landau-Ginzburg free energy expansions \cite{Ishibashi1976,Cao1990,Rychetsky1993,Rychetsky1994,Marton2010}, microscopic theory \cite{Wojdel2014,Kvasov2016,Jiang2017,Valdez2016,Stengel2017}, etc. 
Usually these methods are applied independently of each other. In Landau-Ginzburg theory \cite{Toledano1987} the concept of an order parameter was exploited most successfully in hundreds of cases to describe bulk properties of crystals near structural phase transitions. It turned out to be also very useful for the description of domain wall properties. In Landau theory the domain states (DS) are represented by points in order parameter space, whereas the domain wall is described by a continuous trajectory in the  the order parameter space, connecting the values of the corresponding DSs.
One of the most successful methods for the description of polarization profiles in domain walls uses a modification of the Landau-Ginzburg free energy expansion by adding gradient coupling terms, e.g. flexo-electric couplings \cite{Morozovska2012,Zubko2013,SaljeLi2016} between the strain gradient and the polarization as well as biquadratic terms between OP and polarization. 
   
In virtue of the Curie principle, the presence or absence of polarity within a domain boundary straightforwardly follows from the domain boundary symmetry. Therefore, the central problem consists in determination of the domain boundary symmetry. Obviously, the pure symmetry arguments can only determine the maximal possible symmetry of the domain boundary, which compatible with a given pair of domain states, given crystallographic orientation of the boundary, and possibly also with its exact location in the lattice.
Ensemble of  symmetry operations satisfying simultaneously all these conditions form
the key object of the theory, a layer group $T_{ij}$. This symmetry group $T_{ij}$  can be determined by a detailed inspection of the correspondence between symmetry operations of the parent and child space groups of the crystal structures using a well established systematical abstract group-theoretical approach \cite{Janovec1976,Janovec1981,Janovec2006,Janovec1997,Janovec1989,Kopsky2008,Janovec2004,Janovec2011,Privratska1997,Privratska2000}.

This formal procedure can be apparently circumvented by a more simple approach, based on the symmetry of the averaged order parameter only \cite{Toledano2014}.
The order parameter is represented in a d-dimensional vector space $\textbf{V}=(\eta_1,...,\eta_d)$, depending on the dimension d of the irreducible active representation $\tau_{\alpha \beta}(g) ~(\alpha, \beta=1,...,d)$. 
Two adjacent domains are then represented by two vectors $\textbf{V}_i = (\eta_1^{(i)},..,\eta_d^{(i)})$ and $\textbf{V}_j = (\eta_1^{(j)},..,\eta_d^{(j)})$.    
The main conjecture is that the  symmetry group of a domain wall is at most the maximal isotropy subgroup $A_{ij}$, which leaves $\textbf{V}_i+\textbf{V}_j$ intact. We state equivalently that $A_{ij}$ preserves the arithmetic avarage of the order parameters,
$\langle \textbf{V} \rangle=(\textbf{V}_i+\textbf{V}_j)/2$. 
However, this $A_{ij}$ symmetry group frequently provides only a weak restriction on the properties of a given boundary, because the construction of $A_{ij}$ completely ignores the symmetry-breaking impact of the crystallographic orientation of the domain boundary, the information about the exact location of the boundary in the lattice, and the 2d translational symmetry  of the domain boundary as well. Moreover, there is a possibility of confusion about the role of $A_{ij}$ and $T_{ij}$ groups, which could lead to misinterpretations of  theoretical predictions. For example, nonpolar domain wall symmetries have been indicated \cite{Toledano2014} for
 ferroelastic domain walls of LaAlO$_3$ and SrTiO$_3$, what appears to be in a flagrant contradiction with the recent experimental findings supporting their polarity \cite{Yokota2018,Frenkel2017}, even though long time ago the rigorous theoretical arguments based on layer group methods already disclosed that a lower, polar symmetry is unavoidable there\cite{Janovec1999,Janovec2006}.

The aim of the present work is revise the possibility to assess the presence of polarity in domain boundaries from the point of view of symmetry theory. For this purpose, we show how the procedure of finding the  $T_{ij}$ layer group can be facilitated by considerations about the OP symmetry. 
We  emphasize the fundamental differences among the symmetry group of the domain boundary $T_{ij}$, the symmetry group $A_{ij}$ of the averaged order parameter, and the symmetry groups of  $F_{ij}$  and $J_{ij}$ of the ordered and unordered domain state pairs, respectively. Moreover, we argue that the layer groups $T_{ij}$ allow to verify easily the completeness of Ginzburg-Landau models applied to determine domain wall profiles. General results are illustrated by an explicit analysis for several orientational and translational domain boundaries in real materials.

The paper is organized as follows.
In section \ref{sec:KSCN} we review the main information concerning the phase transition and domain states of KCSN. The symmetry of the intermediate states on the domain wall paths in the OP space is briefly introduced in section \ref{sec:Epikernel}.  
In \ref{sec:LGA} we show how the symmetry of ordered and unordered domain pairs, $F_{ij}$ and $J_{ij}$, is efficiently calculated in order parameter space.   
Section \ref{sec:DT} describes  how the layer group method complemented with order-parameter symmetry allows to obtain the symmetry groups $T_{ij}$ for  selected domain boundaries of KSCN and lacunar spinels. In section \ref{sec:V} we use the resulting layer group-symmetries of domain boundaries $T_{ij}$ to determine symmetry aspects of domain wall trajectories in OP spaces.  Requirements for the adequate quantitative Landau-Ginzburg calculations of domain wall properties are discussed in Section\, VII. Last two sections are devoted to the discussion of the polarity of the domain boundaries and to the general conclusion, respectively.

\section{Phase Transition in KSCN and its Domain States} \label{sec:KSCN}

\begin{figure*}
\centering
\includegraphics[scale=0.5]{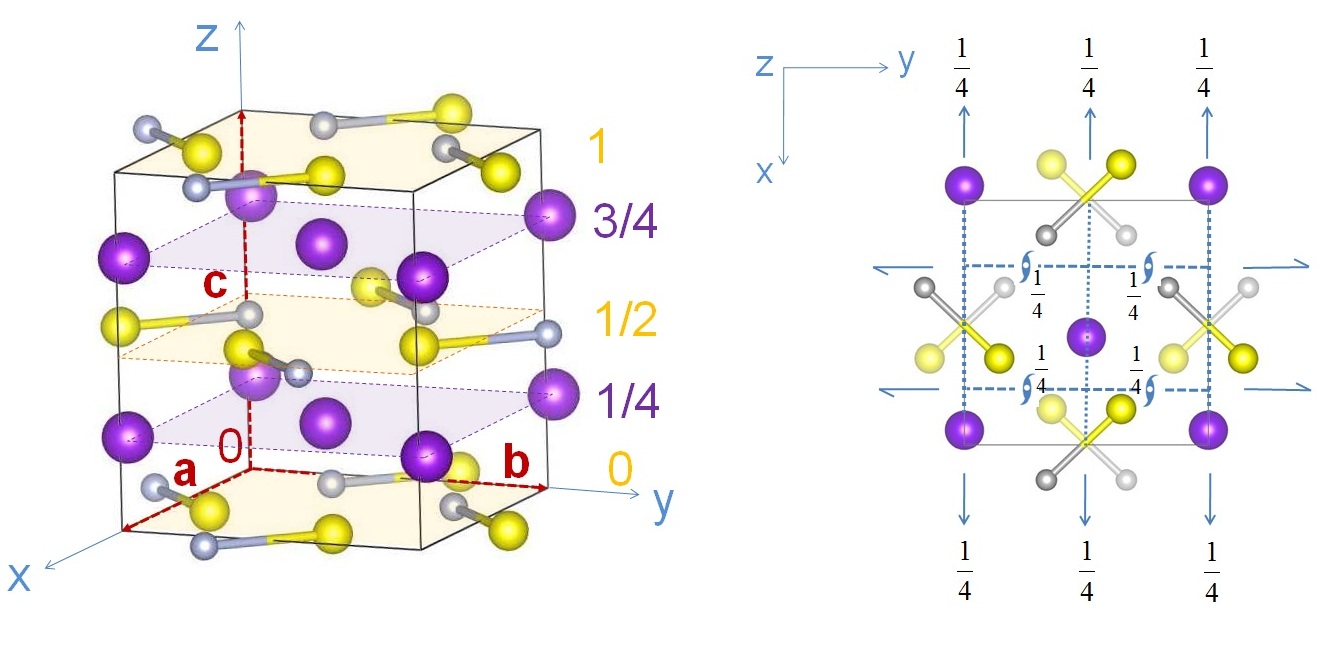}  
\caption{(Left) Three dimensional arrangement of atoms in the orthorhombic $Pbcm$ structure of KSCN. 
The vectors $\textbf{a},\textbf{b},\textbf{c}$ of the primitive unit cell are marked by red dashed arrows. Purple = K atoms, yellow = S, grey = N. C atoms are omitted for clarity in the figures. (Right) c-projection of the structure including symmetry elements. K atoms at $z=\frac{1}{4}c$ levels and S, N atoms at $z=0, \frac{1}{2}c$ levels, respectively.}
\label{fig:KSCN structure}
\end{figure*}

\begin{figure*}
\centering
\includegraphics[scale=0.3]{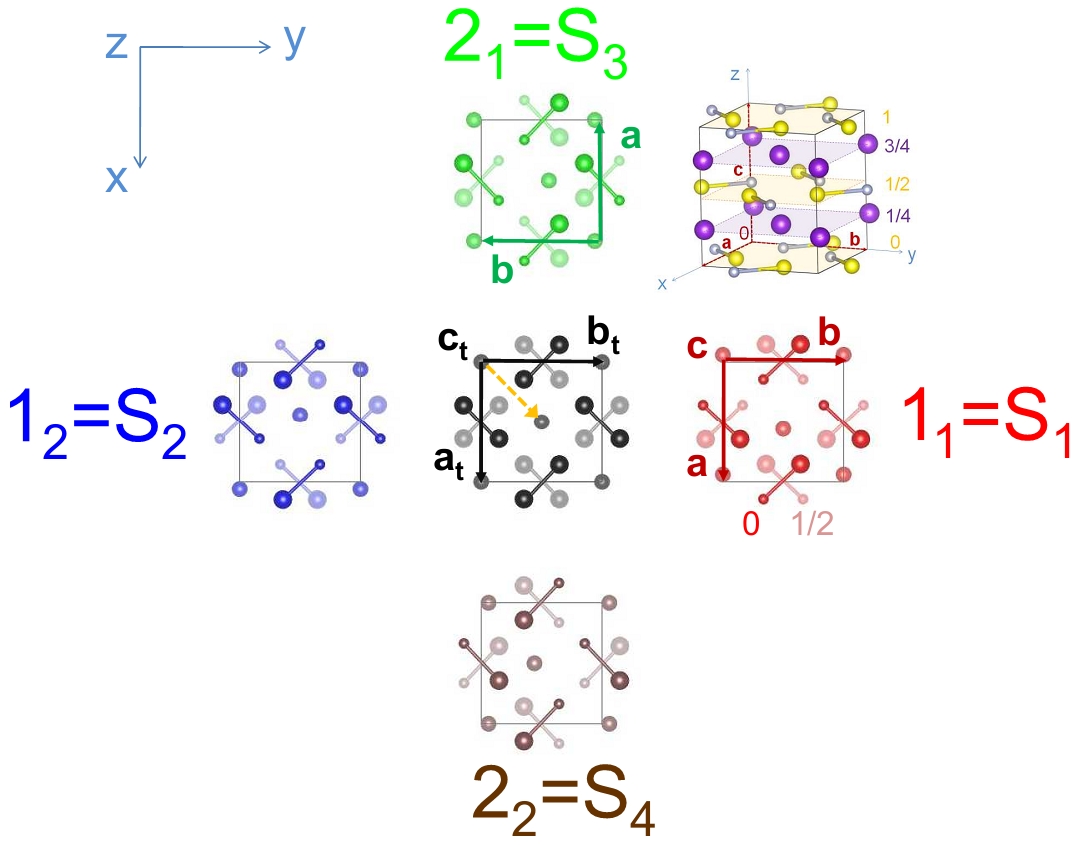}
\caption{4 domain states of the $Pbcm$ structure of KSCN. The conventional unit cell with vectors $\textbf{a}_t=(a_t,0,0)$, $\textbf{b}_t=(0,a_t,0)$ and $\textbf{c}_t=(0,0,c_t)$ of the high temperature tetragonal structure $I4/mcm$ is depicted in the centre. The centring translation $(\frac{1}{2},\frac{1}{2},\frac{1}{2})$ is indicated by a dashed yellow arrow. It is lost at the phase transition to the $Pbcm$-phase $(1_1)$, leading to a doubling of the unit cell below $T_c$.}
\label{fig:DSs}
\end{figure*}

KSCN crystals undergo a structural (order-disorder) phase transition at $T_c$= 415 K. The high temperature phase has a tetragonal body-centred structure with 2 formula units in the primitive unit cell with space group \cite{Yamada1963} $G_0 = I4/mcm ~ (D_{4h}^{18})$, where the $SCN^-$ molecular ions are orientationally (head-tail) disordered. The lattice constants of the conventional unit cell $\textbf{a}_t=(a_t,0,0)$,  $\textbf{b}_t=(0,a_t,0)$, $\textbf{c}_t=(0,0,c_t)$ are $a_t = 6.740$~\AA, $c_t = 7.832$~\AA.\\  
Below $T_c$ the $SCN^-$ molecular ions order in an alternating arrangement, resulting in the loss of the centering translation $(\frac{1}{2},\frac{1}{2},\frac{1}{2})$ which leads to the orthorhombic space group \cite{Yamamoto1987} $F = Pbcm ~ (D_{2h}^{11})$ with 4 formula units in the unit cell. The lattice constants of the orthorhombic unit cell $\textbf{a} = (a,0,0)$, $\textbf{b} = (0,b,0)$, $\textbf{c} = (0,0,c)$ are $a = 6.691$~\AA, $b = 6.676$~\AA, $c = 7.606$~\AA. 
The arrangement of atoms in the orthorhombic phase is shown in Fig.\,\ref{fig:KSCN structure}.

\begin{table*}
\caption{Two dimensional active irreducible representation \cite{Kovalev1965} $\tau_9$  of the tetragonal space group $D_{4h}^{18}$ with order parameter components $(\eta_1,\eta_2)$. The colours denote symmetry operations $g \in D_{4h}^{18}$, which transform $S_1$ into $S_j$, using the colour code of Fig.\,\ref{fig:DSs}. The translational part consists of $\textbf{T} = n\textbf{a} +m\textbf{b} +l\textbf{c}$, $n,m,l \in \mathbb{Z}$ and $e^{i\textbf{k}_c\textbf{T}} = 1$. The components of the polarization vector $\textbf{P}=(P_x,P_y,P_z)$ transform according to the vector representation $V_{ij}(h)$ (i,j=1,2,3), of the corresponding point group elements $h \in D_{4h}$.}
{\begin{tabular}{ccccccccc} \hline \hline \\
\color{red}$(1/000)$ & \color{OliveGreen}$(4_z/000)$ & \color{blue}$(2_z/000)$ & \color{RawSienna}$(4_z^3/000)$ & \color{red}$(2_x/00\frac{1}{2})$ & \color{RawSienna}$(2_{\bar{x}}y)/00\frac{1}{2})$ & \color{blue}$(2_y/00\frac{1}{2})$ & \color{OliveGreen}$(2_{xy}/00\frac{1}{2})$ & OP \\ \hline
$ \begin{pmatrix} 1 & 0\\ 0 & 1 \end{pmatrix} $ & 
$ \begin{pmatrix} 0 & -1\\ 1 & 0 \end{pmatrix} $ &
$ \begin{pmatrix} -1 & 0\\ 0 & -1 \end{pmatrix} $ &
$ \begin{pmatrix} 0 & 1\\ -1 & 0 \end{pmatrix} $ &
$ \begin{pmatrix} 1 & 0\\ 0 & -1 \end{pmatrix} $ & 
$ \begin{pmatrix} 0 & -1\\ -1 & 0 \end{pmatrix} $ &
$ \begin{pmatrix} -1 & 0\\ 0 & 1 \end{pmatrix} $ & 
$ \begin{pmatrix} 0 & 1\\ 1 & 0 \end{pmatrix} $ & 
$ \left(
\begin{array}{c}
\eta_1\\
\eta_2\\
\end{array}
\right)$ \\ \hline 

$ \begin{pmatrix} 1 & 0 & 0\\ 0 & 1 & 0\\ 0 & 0 & 1 \end{pmatrix} $ & 
$ \begin{pmatrix} 0 & -1 & 0\\ 1 & 0 & 0\\ 0 & 0 & 1 \end{pmatrix} $ &
$ \begin{pmatrix} -1 & 0 & 0\\ 0 & -1 & 0\\ 0 & 0 & 1 \end{pmatrix} $ &
$ \begin{pmatrix} 0 & 1 & 0\\ -1 & 0 & 0\\ 0 & 0 & 1 \end{pmatrix} $ &
$ \begin{pmatrix} 1 & 0 & 0\\ 0 & -1 & 0\\ 0 & 0 & -1 \end{pmatrix} $ & 
$ \begin{pmatrix} 0 & -1 & 0\\ -1 & 0 & 0\\ 0 & 0 & -1 \end{pmatrix} $ &
$ \begin{pmatrix} -1 & 0 & 0\\ 0 & 1 & 0\\ 0 & 0 & -1 \end{pmatrix} $ & 
$ \begin{pmatrix} 0 & 1 & 0\\ 1 & 0 & 0\\ 0 & 0 & -1 \end{pmatrix} $ & 
$ \left(
\begin{array}{c}
P_x\\
P_y\\
P_z\\
\end{array}
\right)$ \\ \hline \hline

\color{blue}$(1/\frac{1}{2}\frac{1}{2}\frac{1}{2})$ & \color{RawSienna}$(4_z/\frac{1}{2}\frac{1}{2}\frac{1}{2})$ & \color{red}$(2_z/\frac{1}{2}\frac{1}{2}\frac{1}{2})$ & \color{OliveGreen}$(4_z^3/\frac{1}{2}\frac{1}{2}\frac{1}{2})$ & \color{blue}$(2_x/\frac{1}{2}\frac{1}{2}0)$ & \color{OliveGreen}$(2_{\bar{x}}y)/\frac{1}{2}\frac{1}{2}0)$ & \color{red}$(2_y/\frac{1}{2}\frac{1}{2}0)$ & 
\color{RawSienna}$(2_{xy}/\frac{1}{2}\frac{1}{2}0)$ & OP \\ \hline
$ \begin{pmatrix} -1 & 0\\ 0 & -1 \end{pmatrix} $ & 
$ \begin{pmatrix} 0 & 1\\ -1 & 0 \end{pmatrix} $ &
$ \begin{pmatrix} 1 & 0\\ 0 & 1 \end{pmatrix} $ &
$ \begin{pmatrix} 0 & -1\\ 1 & 0 \end{pmatrix} $ &
$ \begin{pmatrix} -1 & 0\\ 0 & 1 \end{pmatrix} $ & 
$ \begin{pmatrix} 0 & 1\\ 1 & 0 \end{pmatrix} $ &
$ \begin{pmatrix} 1 & 0\\ 0 & -1 \end{pmatrix} $ & 
$ \begin{pmatrix} 0 & -1\\ -1 & 0 \end{pmatrix} $ & 
$ \left(
\begin{array}{c}
\eta_1\\
\eta_2\\
\end{array}
\right)$ \\ \hline 

$ \begin{pmatrix} 1 & 0 & 0\\ 0 & 1 & 0\\ 0 & 0 & 1 \end{pmatrix} $ & 
$ \begin{pmatrix} 0 & -1 & 0\\ 1 & 0 & 0\\ 0 & 0 & 1 \end{pmatrix} $ &
$ \begin{pmatrix} -1 & 0 & 0\\ 0 & -1 & 0\\ 0 & 0 & 1 \end{pmatrix} $ &
$ \begin{pmatrix} 0 & 1 & 0\\ -1 & 0 & 0\\ 0 & 0 & 1 \end{pmatrix} $ &
$ \begin{pmatrix} 1 & 0 & 0\\ 0 & -1 & 0\\ 0 & 0 & -1 \end{pmatrix} $ & 
$ \begin{pmatrix} 0 & -1 & 0\\ -1 & 0 & 0\\ 0 & 0 & -1 \end{pmatrix} $ &
$ \begin{pmatrix} -1 & 0 & 0\\ 0 & 1 & 0\\ 0 & 0 & -1 \end{pmatrix} $ & 
$ \begin{pmatrix} 0 & 1 & 0\\ 1 & 0 & 0\\ 0 & 0 & -1 \end{pmatrix} $ & 
$ \left(
\begin{array}{c}
P_x\\
P_y\\
P_z\\
\end{array}
\right)$ \\ \hline \hline

\color{blue}$(\bar{1}/000)$ & \color{RawSienna}$(\bar{4}_z/000)$ & \color{red}$(m_z/000)$ & \color{OliveGreen}$(\bar{4}_z^3/000)$ & \color{blue}$(m_x/00\frac{1}{2})$ & \color{OliveGreen}$(m_{\bar{x}}y)/00\frac{1}{2})$ & \color{red}$(m_y/00\frac{1}{2})$ & \color{RawSienna}$(m_{xy}/00\frac{1}{2})$ & OP \\ \hline
$ \begin{pmatrix} -1 & 0\\ 0 & -1 \end{pmatrix} $ & 
$ \begin{pmatrix} 0 & 1\\ -1 & 0 \end{pmatrix} $ &
$ \begin{pmatrix} 1 & 0\\ 0 & 1 \end{pmatrix} $ &
$ \begin{pmatrix} 0 & -1\\ 1 & 0 \end{pmatrix} $ &
$ \begin{pmatrix} -1 & 0\\ 0 & 1 \end{pmatrix} $ & 
$ \begin{pmatrix} 0 & 1\\ 1 & 0 \end{pmatrix} $ &
$ \begin{pmatrix} 1 & 0\\ 0 & -1 \end{pmatrix} $ & 
$ \begin{pmatrix} 0 & -1\\ -1 & 0 \end{pmatrix} $ & 
$ \left(
\begin{array}{c}
\eta_1\\
\eta_2\\
\end{array}
\right)$ \\ \hline 

$ \begin{pmatrix} -1 & 0 & 0\\ 0 & -1 & 0\\ 0 & 0 & -1 \end{pmatrix} $ & 
$ \begin{pmatrix} 0 & 1 & 0\\ -1 & 0 & 0\\ 0 & 0 & -1 \end{pmatrix} $ &
$ \begin{pmatrix} 1 & 0 & 0\\ 0 & 1 & 0\\ 0 & 0 & -1 \end{pmatrix} $ &
$ \begin{pmatrix} 0 & -1 & 0\\ 1 & 0 & 0\\ 0 & 0 & -1 \end{pmatrix} $ &
$ \begin{pmatrix} -1 & 0 & 0\\ 0 & 1 & 0\\ 0 & 0 & 1 \end{pmatrix} $ & 
$ \begin{pmatrix} 0 & 1 & 0\\ 1 & 0 & 0\\ 0 & 0 & 1 \end{pmatrix} $ &
$ \begin{pmatrix} 1 & 0 & 0\\ 0 & -1 & 0\\ 0 & 0 & 1 \end{pmatrix} $ & 
$ \begin{pmatrix} 0 & -1 & 0\\ -1 & 0 & 0\\ 0 & 0 & 1 \end{pmatrix} $ & 
$ \left(
\begin{array}{c}
P_x\\
P_y\\
P_z\\
\end{array}
\right)$ \\ \hline \hline

\color{red}$(\bar{1}/\frac{1}{2}\frac{1}{2}\frac{1}{2})$ & \color{OliveGreen}$(\bar{4}_z/\frac{1}{2}\frac{1}{2}\frac{1}{2})$ & \color{blue}$(m_z/\frac{1}{2}\frac{1}{2}\frac{1}{2})$ & \color{RawSienna}$(\bar{4}_z^3/\frac{1}{2}\frac{1}{2}\frac{1}{2})$ & \color{red}$(m_x/\frac{1}{2}\frac{1}{2}0)$ & \color{RawSienna}$(m_{\bar{x}}y)/\frac{1}{2}\frac{1}{2}0)$ & \color{blue}$(m_y/\frac{1}{2}\frac{1}{2}0)$ & \color{OliveGreen}$(m_{xy}/\frac{1}{2}\frac{1}{2}0)$ & OP \\ \hline
$ \begin{pmatrix} 1 & 0\\ 0 & 1 \end{pmatrix} $ & 
$ \begin{pmatrix} 0 & -1\\ 1 & 0 \end{pmatrix} $ &
$ \begin{pmatrix} -1 & 0\\ 0 & -1 \end{pmatrix} $ &
$ \begin{pmatrix} 0 & 1\\ -1 & 0 \end{pmatrix} $ &
$ \begin{pmatrix} 1 & 0\\ 0 & -1 \end{pmatrix} $ & 
$ \begin{pmatrix} 0 & -1\\ -1 & 0 \end{pmatrix} $ &
$ \begin{pmatrix} -1 & 0\\ 0 & 1 \end{pmatrix} $ & 
$ \begin{pmatrix} 0 & 1\\ 1 & 0 \end{pmatrix} $ & 
$ \left(
\begin{array}{c}
\eta_1\\
\eta_2\\
\end{array}
\right)$ \\ \hline

$ \begin{pmatrix} -1 & 0 & 0\\ 0 & -1 & 0\\ 0 & 0 & -1 \end{pmatrix} $ & 
$ \begin{pmatrix} 0 & 1 & 0\\ -1 & 0 & 0\\ 0 & 0 & -1 \end{pmatrix} $ &
$ \begin{pmatrix} 1 & 0 & 0\\ 0 & 1 & 0\\ 0 & 0 & -1 \end{pmatrix} $ &
$ \begin{pmatrix} 0 & -1 & 0\\ 1 & 0 & 0\\ 0 & 0 & -1 \end{pmatrix} $ &
$ \begin{pmatrix} -1 & 0 & 0\\ 0 & 1 & 0\\ 0 & 0 & 1 \end{pmatrix} $ & 
$ \begin{pmatrix} 0 & 1 & 0\\ 1 & 0 & 0\\ 0 & 0 & 1 \end{pmatrix} $ &
$ \begin{pmatrix} 1 & 0 & 0\\ 0 & -1 & 0\\ 0 & 0 & 1 \end{pmatrix} $ & 
$ \begin{pmatrix} 0 & -1 & 0\\ -1 & 0 & 0\\ 0 & 0 & 1 \end{pmatrix} $ & 
$ \left(
\begin{array}{c}
P_x\\
P_y\\
P_z\\
\end{array}
\right)$ \\ \hline \hline

\end{tabular}}
\label{tab:tableI}
\end{table*}

Due to the symmetry reduction at the phase transition, the number \cite{Janovec1989} of domain states (DSs) $n=4$. We denote them as $S_1 = 1_1$, $S_2 = 1_2$, $S_3 = 2_1$, $S_4 = 2_2$, where the main index numbers the two \textit{orientational} DSs $1_1, 2_1$ and the subindex distinguishes the two different \textit{translational} DSs $1_1, 1_2$. The 4 possible DS's are shown in Fig.\,\ref{fig:DSs}.
All operations that transform $S_1$ into $S_j ~ (j=1,..,4)$ are marked in Table \ref{tab:tableI} by colours using the colour code of Fig. \ref{fig:DSs}.

\begin{figure}
\centering
\includegraphics[scale=0.5]{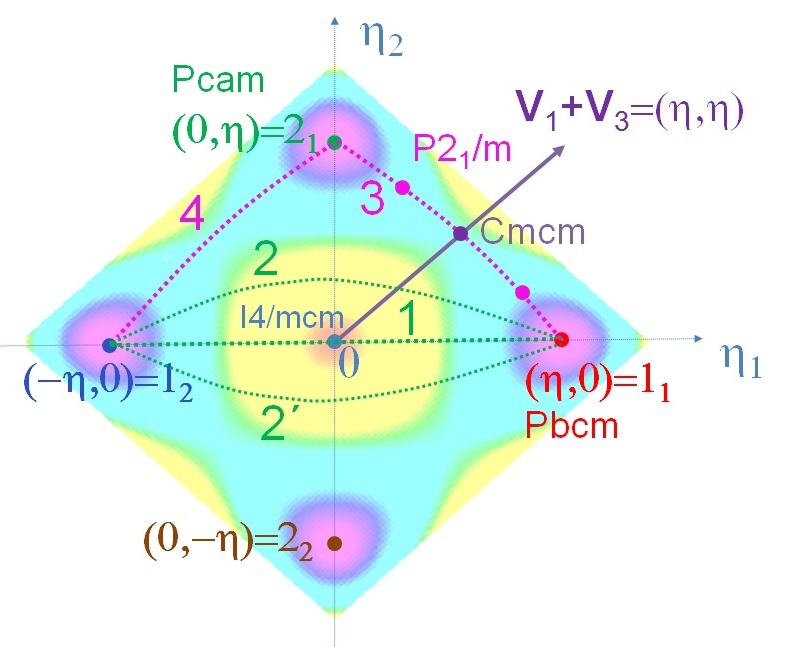}  
\caption{Sketch of the Landau free energy landscape in the OP space $(\eta_1,\eta_2)$ of KSCN, representation of homogeneous domain states (points) and transition pathways (dotted lines) between them. 1 = straight path, also called linear anti-phase boundary (LAPB), 2 and 2' are side paths, also called rotational anti-phase boundaries (RAPBs), 3 = ferroelastic domain boundary. Path 4 describes another ferroelastic boundary.    
The epikernel symmetries at the points or segments of the transition pathways \cite{Schranz1995} are also depicted. Generally they do not correspond to the correct local structures, since the rate of structural changes is usually not small, especially close and at the domain wall centres.}
\label{fig:OP space}
\end{figure}

For further considerations, we shortly review some results from the Landau theory of KSCN \cite{Schranz1989,Schranz1994}. 
The phase transition of KSCN \cite{Schranz1989} from $I4/mcm (D_{4h}^{18})$ to $Pbcm (D_{2h}^{11})$ occurs at the critical wavevector $\textbf{k}_c = (00\frac{2\pi}{c})$.
It describes a wave which has its period equal to the $c$-axes, but the centring translation $(\frac{1}{2} \frac{1}{2} \frac{1}{2})$ of the tetragonal $D_{4h}^{18}$ phase is lost.  
The PT is described by the two dimensional irreducible representation $\tau_9$ (Table \ref{tab:tableI}) \cite{Kovalev1965} with the OP components $(\eta_1,\eta_2)$. 
The symmetry $Pbcm (D_{2h}^{11})$ requires a minimum of the free energy at $(\eta,0)$. In OP space the 4 (homogeneous) domain states can then be located as points at $(\eta,0) \equiv S_1=1_1$, $(-\eta,0) \equiv S_2=1_2$, $(0,\eta) \equiv S_3=2_1$ and $(0,-\eta) \equiv S_4=2_2$ (Fig.\,\ref{fig:OP space}).  

\section{ Domain wall paths and intermediate states in OP space}\label{sec:Epikernel}

A smoothly varying structure of a domain wall bridging $S_i$ and $S_j$ can be associated with a path in OP-space which connects $S_i$ and $S_j$.  
To each position $\xi\textbf{n}$ (\textbf{n} is the normal to the domain wall) in real space there corresponds a point $(\eta_1(\xi),\eta_2(\xi))$ on the path \cite{Janovec1981}, which describes the local structure of the domain wall. 
The rate of the structure change along the path can be expressed as $h(\xi)=\sqrt{\left(\frac{\partial \eta_1}{\partial \xi}\right)^2+\left(\frac{\partial \eta_2}{\partial \xi}\right)^2}$. 
Small values of $h$ imply nearly homogeneous regions as it is the case for $\xi \rightarrow \pm \infty$. 
Close to the domain wall center ($\xi \rightarrow 0$) $h$ becomes usually very large. If $h$ is small, the local symmetry at position $\xi$ in a domain wall can be described by a three dimensional space group $E(\eta_1(\xi),\eta_2(\xi))$, which represents the epikernels \cite{Ascher1977} of the representation inducing the phase transition. However, at positions $\xi$ with high $h$ - i.e. close to the domain wall center - the local symmetry should be described by a layer group  with 2d periodicity, which of course is different from the 3 dimensional space group $E(\eta_1(\xi),\eta_2(\xi))$. 
Not taking this into account can lead to ambiguous results, as we will show on the following examples.

\subsection{Ferroelastic domain boundary paths of KSCN} \label{subsec:IIA}

For a $(110)$- or $(1\bar{1}0)$-oriented domain wall (compatible) between $1_1$ and $2_1$ or $1_2$ and $2_1$ (paths 3 or 4 in Fig.~\ref{fig:OP space}) the OP components $(\eta_1(\xi),\eta_2(\xi))$ vary between $(\eta,0)$ and $(0,\eta)$ (or $(-\eta,0)$ and $(0,\eta)$) via $(\eta,\eta)$ (or $(-\eta,\eta)$) in the domain wall centre ($\xi =0$). 
If we use Table\,\ref{tab:tableI} to find out, which symmetry operations leave $(\eta_1(\xi),\eta_2(\xi))$ unchanged along the path  $-\infty \leq \xi \leq \infty$, we obtain: $Pbcm (\xi = -\infty) \rightarrow P2_1/m (-\infty < \xi < 0) \rightarrow Cmcm (\xi = 0) \rightarrow P2_1/m (0 < \xi < \infty) \rightarrow Pcam (\xi = \infty)$.
Thus, the symmetry of the domain wall centre $(Cmcm)$ coincides with the symmetry $A_{13}$ of $\textbf{V}_1 + \textbf{V}_3 = (\eta,0)+(0,\eta) = (\eta,\eta)$ of Ref.\,\onlinecite{Toledano2014}. 
 
All these groups are non-polar, in contrast to the general statement, saying that all mechanically compatible ferroelastic domain walls must be polar \cite{Janovec1999}. The reason for this overestimation of symmetry is, that in the above approximations a 3-dimensional domain wall structure is assumed, while a planar domain wall has only 2-dimensional periodicity within this plane.

\subsection{Anti-phase boundary paths of KSCN}  \label{subsec:IIB}

There are three different ways (paths 1, 2 or 2' in Fig.\,\ref{fig:OP space}) to connect $1_1$ and $1_2$ along a given path $\xi$ via an translational anti-phase boundary.\\
Path 1 leads - using table \ref{tab:tableI} - to the following sequence of epikernel symmetries:
$Pbcm (- \infty \leq \xi < 0)$, $I4/mcm (\xi =0)$, $Pbcm (0  < \xi \leq \infty)$.\\ 
Paths 2 and 2' lead to:
$Pbcm (\xi = - \infty) \rightarrow P2_1/m (- \infty < \xi < 0) \rightarrow Pcam (\xi =0) \rightarrow P2_1/m (0  < \xi < \infty) \rightarrow Pbcm (\xi = \infty)$. Let us not that the average order parameter symmetry \cite{Toledano2014} $A_{12}$ of $(\textbf{V}_1 + \textbf{V}_2) = ((\eta,0)+(-\eta,0)) = (0,0) $ yields  $I4/mcm$, independently on the path between $1_1$ and $1_2$.
It should be noted, that also for the translational anti-phase boundaries both approaches yield non-polar groups, in contrast to our previous findings \cite{Janovec1989}, where we have used layer groups to analyse the domain wall symmetries. Also here, the obtained symmetries are too high, since the two dimensional character of the domain wall is not properly taken into account.
The change of symmetry within a domain wall along the path $\xi$ is taken fully into account by the layer group method \cite{Janovec2006}, which is widely used in the next sections.

\section{Symmetry of domain pairs } \label{sec:LGA}

 Previously, a detailed symmetry analysis of domain pairs and  boundaries in KSCN have been performed in Refs.\cite{Janovec1989, Janovec2006}. Here we use the same settings and notations, but at some steps we used advantage of the known irrep of the order parameter (OP). Moreover - as we show below - working in OP space helps a lot to connect to Landau-Ginzburg theory, i.e. to find the most important coupling terms, which are needed to describe the (functional) properties of the corresponding domain walls.  

To find the symmetry of a domain wall between domain states $S_i$ and $S_j$ with the symmetry groups $F_i$ and $F_j$, one usually starts with the symmetry analysis of a corresponding  \textit{domain  pair} (DP). A DP represents an intermediate step between domain states (DS) and domain walls and can be visualized as two overlapping structures $S_i$ and $S_j$, which exist independently of each other, both filling the entire space.  
It can be treated either as an \textit{unordered  domain  pair}
\begin{equation}
\label{eq:unorderdered pair}
\{S_i,  S_j \} = \{ S_j,  S_i \}   
\end{equation}
or an \textit{ordered} domain pair
\begin{equation}
\label{eq:ordered pair}
( S_i,  S_j) \neq ( S_j,  S_i)~, 
\end{equation}
where $( S_j,  S_i)$ is a \textit{transposed} domain pair of $( S_i,  S_j)$.

Operations $f \in G_0$ that leave both $S_i$ and $S_j$ unchanged are operations common to $F_i$ and $F_j$. They form a group $F_{ij}$
\begin{equation}
\label{eq:F12}
F_{ij} = F_{i} \cap  F_{j} ~.
\end{equation}
The group $F_{ij}$ is thus the symmetry group of an ordered domain pair $( S_i,  S_j)$.

The symmetry group $J_{ij}$ of an unordered domain pair $\{S_i,  S_j \}$ consists of the group $J_{ij}'=F_{ij}$ and in addition, it contains all  \textit{transposing} operations $\hat{\jmath}_{ij} \in G_0$ which transform the ordered pair $( S_i,  S_j)$ into the transposed domain pair $( S_j,  S_i)$.
All transposing operations are contained in the left coset $J_{ij}''=\hat{\jmath}_{ij}F_{ij}$.  Thus the \textit{symmetry group} $J_{ij}$  of the \textit{unordered}  domain pair  $\{S_i,S_j\}$ is equal to   
\begin{equation}
\label{eq:Jij}
J_{ij} = J_{ij}' \cup J_{ij}'' = F_{ij} \cup \hat{\jmath}_{ij}F_{ij}  ~.
\end{equation}  
The group $J_{ij}$ can be treated as a dichromatic (e.g. black and white) group \cite{Bradley1972}. If one colours the domain states, say $S_i$ black  and $ S_j$ white, then operations $f\in J_{ij}'=F_{ij}$ (without caret) can be treated as colour-preserving operations whereas operations with caret $\hat{f} \in \hat{\jmath}_{ij} F_{ij} = J_{ij}''$ as colour-changing  ones. 

Let us apply now this procedure to find the symmetry groups $J_{ij}$ of unordered DPs in KSCN using order parameter symmetries. In the present case there are six DPs, denoted as $\{1_1,1_2\}$, $\{1_1,2_1\}$, $\{1_1,2_2\}$, $\{1_2,2_1\}$, $\{1_2,2_2\}$, $\{2_1,2_2\}$. According to  table~IV of Ref.\,\onlinecite{Janovec1989}, there are two sets of symmetrically inequivalent DPs. Out of these we will consider only the two inequivalent DPs $\{ S_1, S_2\}= \{ 1_1, 1_2\}$ (translational DP) and  $\{ S_1, S_3\}= \{ 1_1, 2_1\}$ (orientational DP).

\subsection{Symmetry of orientational domain pairs of KSCN} \label{IIIA}

\begin{figure}
\centering
\includegraphics[scale=0.7]{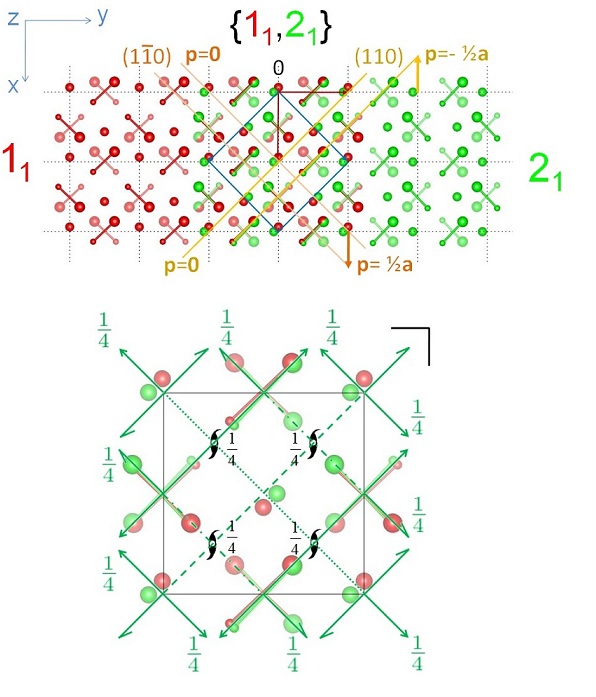} 
\caption{
Top: Orientational DP $\{\color{black}1_1\color{black},\color{black}2_1 \color{black}\}=\{S_1,S_3\}$. The ferroelastic domain walls at positions $\textbf{p}=p\textbf{a}$ for $p=0, \frac{1}{2}, and  -\frac{1}{2}$ are indicated by yellow and orange lines.   
Bottom: Symmetry elements of $J_{13}'$ (that is $(1/000)$, $(2_z/\frac{1}{2} \frac{1}{2}\frac{1}{2})$, $(m_z/000)$, and $(\bar{1}/\frac{1}{2} \frac{1}{2}\frac{1}{2})$, drawn in black)  and of $J_{13}''$ (that is
$ (2_{xy}/00\frac{1}{2})$, $(2_{\bar{x}y}/\frac{1}{2} \frac{1}{2} 0)$,
$(m_{\bar{x}y}/00\frac{1}{2})$, $(m_{\bar{x}y}/00\frac{1}{2})$ and $(m_{xy}/\frac{1}{2} \frac{1}{2} 0)$, drawn in green) forming altogether $J_{13}'\cup J_{13}''= C \widehat{m}_{xy} \widehat{c}_{\bar{x}y} m_z = D_{2h}^{17}$ attached to the structure of the DP. 
The unit cell (blue lines) of $Cmcm$ which is related to $Pbcm$ (deep red) according to $\textbf{a}_o=\textbf{a}-\textbf{b}$, $\textbf{b}_o=\textbf{a}+\textbf{b}$ and $\textbf{c}_o=\textbf{c}$ is also shown. To keep consistency with earlier work, $y$ axis is drawn as horizontal.}
\label{fig:Domain pairs}
\end{figure}

Since a DP corresponds to an overlap of homogeneous DSs, which can be represented as points in OP space, $J_{ij}$ can be conveniently calculated using the OP symmetry in terms of the active irreducible representation (Table~\ref{tab:tableI}).  
To show this, let us first consider the DP $\{ 1_1 , 2_1 \}$ ($=\{S_1, S_3\}$, see Fig.~\ref{fig:DSs}), which we represent in OP-space as $\{(\eta,0),(0,\eta)\}$.
$F_{13}$ consists of all  symmetry elements $f \in I4/mcm$ for which $f\{\color{black}(\eta,0)\color{black},\color{black}(0,\eta)\color{black}\}= \{\color{black}(\eta,0)\color{black},\color{black}(0,\eta)\color{black}\}$, i.e. which leave each of the two DSs unchanged. 
In terms of irreducible representations this condition translates to $\tau(f)\color{black} (\eta, 0) \color{black} = \color{black} (\eta,0) \color{black}$ and $\tau(f)\color{black} (0,\eta) \color{black} = \color{black} (0,\eta) \color{black}$, where $\tau(f)$ is a matrix corresponding to $f$, see Table~\ref{tab:tableI}. According to Table~\ref{tab:tableI} these are the symmetry elements where $\tau_{11}(f)=\tau_{22}(f)=1$ and $\tau_{12}(f)=\tau_{21}(f)=0$. This implies that $ F_{13}$ reads
\begin{equation}
    F_{13} = {\bf T} \{(1/000)~ (2_z/\frac{1}{2} \frac{1}{2}\frac{1}{2}) ~ (m_z/000) ~ (\bar{1}/\frac{1}{2} \frac{1}{2}\frac{1}{2})\} ~,
\end{equation}
with translations (Fig.\,\ref{fig:Domain pairs}) $\textbf{T}=n\textbf{a}_o+m\textbf{b}_o+l\textbf{c}_o$ and $\textbf{a}_o = \textbf{a}-\textbf{b}$, $\textbf{b}_o = \textbf{a} + \textbf{b}$ and $\textbf{c}_o=\textbf{c}$. This is a nonpolar $P2_{1z}/m_z $ space group.

In order to identify the domain state-exchanging symmetry operations $\widehat{j}$ within $ I4/mcm$, it is convenient to exploit again the active irrep of the transition. The searched symmetry operations should fulfill  $ \widehat{j} \{(\eta,0)(0,\eta) \} = \{(0,\eta) (\eta,0)\}$. Therefore, for each such operation the matrix $\tau =\tau(~\widehat{j}~)$ has to satisfy $\tau (\eta, 0) \color{black} = \color{black} (0,\eta) \color{black}$ and $\tau  (0,\eta) \color{black} = \color{black} (\eta,0) \color{black}$, and thus $\tau_{11} =\tau_{22} =0$ and $\tau_{12}=\tau_{21}=1$. 
This is fulfilled for a set of operations $J_{13}''$
\begin{equation}
      {\bf T}\{(2_{xy}/00\frac{1}{2}) ~(2_{\bar{x}y}/\frac{1}{2} \frac{1}{2} 0)~ (m_{\bar{x}y}/00\frac{1}{2}) ~ (m_{xy}/\frac{1}{2} \frac{1}{2} 0) \}~.
\end{equation}
Combining both results yields 
\begin{equation}
    J_{13} = J_{13}' \cup \color{black} J_{13}'' = C \widehat{m}_{xy} \widehat{c}_{\bar{x}y}  m_z = D_{2h}^{17}~, \nonumber
\end{equation}
 again with translations (Fig.\,\ref{fig:Domain pairs}) $\textbf{T}=n\textbf{a}_o+m\textbf{b}_o+l\textbf{c}_o$ and $\textbf{a}_o = \textbf{a}-\textbf{b}$, $\textbf{b}_o = \textbf{a} + \textbf{b}$ and $\textbf{c}_o=\textbf{c}$. 

A graphical picture of the symmetry elements of $J_{13}$ is given in Fig.\,\ref{fig:Domain pairs}. All colour-preserving elements ($ F_{13}$) are marked in black, whereas all colour-changing elements  ($ J_{13}''$) are marked in green. 
It should be noted, that for this DP $\{1_1,2_1\}$ the symmetry $J_{13}$ is coincidentally the same as the $A_{13}$ ($ Cmcm$). Or stating it other way round, for this example the OP-approaches of Ref.~\onlinecite{Toledano2014} and Ref.~\onlinecite{Schranz1995} yield the symmetry of a DP, which generally is higher than the symmetry of the corresponding domain wall, as will be shown below. Resulting domain pair symmetry $J_{13}$ obviously coincides with the result obtained by the original procedure of Ref.\,\onlinecite{Janovec1989}. 

\subsection{Symmetry of translational domain pairs of KSCN} \label{IIIB}

\begin{figure}
\centering
\includegraphics[scale=0.4]{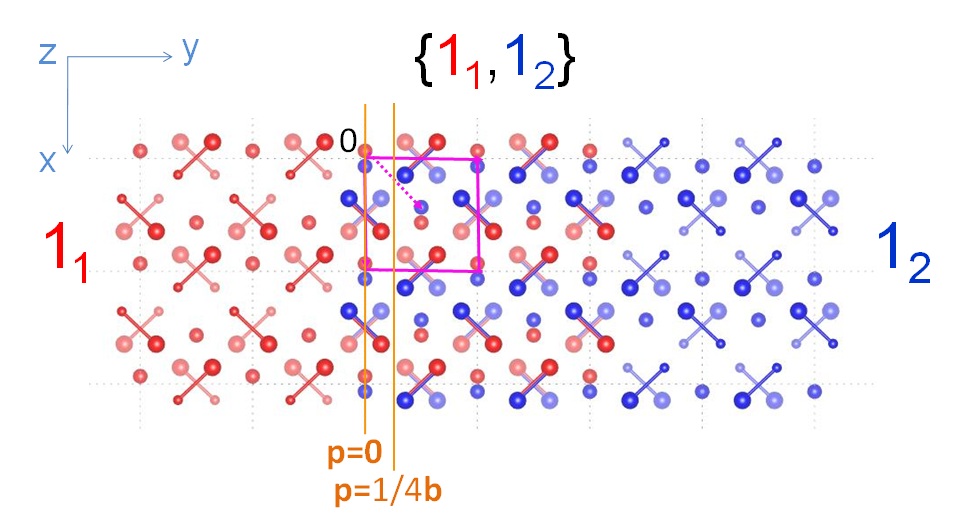}
\caption{Translational DP 
$\{{1_1},{1_2}\} = \{{S_1},{S_2}\}$ 
of the KSCN structure. The unit cell of the group $J_{12}=Ibam=D_{2h}^{26}$ is shown in magenta. Note, that the centring translation (dotted magenta vector) is conserved only, if we omit the colors, i.e. if we treat it as an unordered DP.}
\label{fig:DP}
\end{figure}

The translational DP $\{1_1,1_2\}$ is shown in Fig.~\ref{fig:DP}. 
The color preserving operations of $F_{12}=F_1 \cap F_2$ are those, which leave both, $(\eta,0)$ and $(-\eta,0)$ unchanged. According to table~\ref{tab:tableI} these are the elements marked in red, i.e. $F_{12}= Pbcm$. The color changing operations are those with $\tau_{11}=-1$. They are marked in table~\ref{tab:tableI} in blue color. 
Thus the symmetry elements that leave the DP $\{1_1,1_2\}$ invariant form the space group 
\begin{equation}
J_{12} = Pbcm + \textcolor{blue} {\widehat{2}_z} Pbcm = I \widehat{b}_x \widehat{a}_y m_z = D_{2h}^{26} ~.
\end{equation}
 It consists of the union of elements marked in red and blue in table~\ref{tab:tableI} with $T=n\textbf{a}+m\textbf{b}+l\textbf{c}$.
For this DP the symmetry group \cite{Toledano2014} $G(\textbf{V}_1 + \textbf{V}_2) = I4/mcm$  (see \ref{subsec:IIB}) is even higher than the symmetry ($Ibam$) of the unordered DP.

\section{Symmetry of domain boundaries} \label{sec:DT}

In a next step we complement the layer group approach \cite{Janovec1981} by OP symmetry and calculate the symmetry of domain boundaries  \cite{Janovec1989} using layer groups and irreps. 

Sometimes it is convenient to introduce the concept of domain twin, which consists of two semi-infinite domains which meet along a planar transitional layer region, called domain wall or domain boundary. With this definition, symmetry of both objects  (planar domain twin and planar domain boundary) is the same.
A position of planar domain boundary can be defined by a normal \textbf{n} to the boundary plane and one selected position vector \textbf{p} within this plane.  The vector \textbf{n} defines also the sidedness of the arrangement, i.e. the side of  the first domain state with respect to \textbf{n}. Both vectors are typically defined with respect to the parent or the child crystal lattice, \textbf{p} is understood as a position with respect to the origin of the crystallographic cell.  Sometimes it is convenient to select the position vector of the boundary as its intercept $p$ with respect to the origin of the selected unit cell, $\textbf{p} =p{\bf n}$. 
A convenient symbol of a domain boundary is then $(S_i | \textbf{n},\textbf{p} | S_j)$ or $(S_i | \textbf{n},p | S_j)$. 
All $g \in D_{4h}^{18}$ that leave the domain boundary invariant, i.e. for which
\begin{equation}
\label{eq:layer group}
g(S_i|\textbf{n},\textbf{p}|S_j) = (gS_i|g\textbf{n},g\textbf{p}|gS_j) = (S_i|\textbf{n},\textbf{p}|S_j)
\end{equation} 
holds, form a layer group (space group with 2 d periodicity) $T_{ij}$ \cite{Vainstein1981}, which determines the symmetry of the domain boundary.
Sometimes dependence of the layer group on $\textbf{n}$ and $\textbf{p}$ will be shown explicitly as $T_{ij}(\textbf{n},\textbf{p})$. 

Generally the group $T_{ij}$ consists of two parts
\begin{equation}
\label{eq:layer group decomposition}
T_{ij} = T_{ij}' \cup T_{ij}'' = \bar{F}_{ij} \cup \widehat{\underbar{t}}_{ij} \bar{F}_{ij} ~, 
\end{equation} 
where $T_{ij}'=\bar{F}_{ij} $  consists of all operations in $F_{ij}$ that leave $S_i,S_j,\textbf{n},\textbf{p}$ invariant while $T_{ij}'' =\widehat{\underbar{t}}_{ij} \bar{F}_{ij}$ consists of all operations of $ J_{ij}'' $  that  simultaneously exchange $S_i$ and $S_j$ and transform $\textbf{n}$ into $-\textbf{n}$ (the latter property is maked by underlining the symbols of such operations). 
If $T_{ij}'' = 0$, then $T_{ij} = T_{ij}'$, and the domain boundary is denoted an \textit{asymmetric} domain boundary. In the opposite case, when $T_{ij}'' \neq 0$, the domain boundary is called a \textit{symmetric} domain boundary. These symmetry properties of domain boundaries are important in the context of domain wall properties.
In the following we show, how the symmetry elements of $T_{ij}', T_{ij}''$ and $T_{ij}$  can be systematically calculated by inspecting how the order parameter components transform under the action of a given symmetry element, however taking into account how \textbf{n} transforms at the position \textbf{p}.

\subsection{Ferroelastic domain boundaries of KSCN} \label{IVA}

It is known, that for the present symmetry reduction $I4/mcm \rightarrow Pbcm$ there exist two elastically compatible domain wall orientations, i.e. $\textbf{n} =(1,1,0)$ and $(1,\bar{1},0)$ for orientational (ferroelastic) domains. 
First we calculate the symmetry of an orientational (ferroelastic) domain boundary 
$(1_1|(1,1,0),0|2_1)$.

By inspecting Fig.~\ref{fig:Domain pairs}, we find those operations of $ F_{13} = P2_1/m_z$, which at $p = 0$ leave $\textbf{n} = (1,1,0)$ invariant. Invariant of $\textbf{n}$ at $\textbf{p}$ means, that those operations should neither change the orientation of $\textbf{n}$, nor shift the domain wall from its position \textbf{p}.  
These elements form the layer group $T_{13}' = \textbf{T} \{(1/000)~(m_z/000)\} \equiv pm_z$. Since only shifts within the domain wall-plane are allowed, one obtains
$\textbf{T} = n(\textbf{a}-\textbf{b})+m\textbf{c} ~ (n,m \in \mathbb{Z})$.
There are no position-preserving operations within
 $ J_{13}''$, which at $\textbf{p} = \textbf{0}$ change $\textbf{n} \rightarrow -\textbf{n}$, so that 
\begin{equation}
   T_{13} = T_{13}'= \textbf{T} \{(1/000)~(m_z/000) \} ~. 
\end{equation}
These symmetry elements form the layer group $T_{13} \equiv p  m_z$ (note, that layer group symmetries are marked by small letters in front of the symbol). 
It is obvious, that this layer group $T_{13}$ of the domain boundary is polar and thus allows for a polarization component $P_{[1\bar{1}0]}$ in the centre of the domain wall.

\begin{table} 
\caption{Layer group symmetry of ferroelastic  domain boundaries of KSCN.}
{\begin{tabular}{lccc} \hline \hline \\
Domain boundary & Position p &  Layer group \\ \hline 
$(1_1|~ (110),p ~|2_1)$ & $0$  & $T_{13}$ = $pm_z$ \\ 
             &   &  $\bar{F}_{13}$ = $pm_z$ \\ 
             & $-\frac{1}{2}$  & $T_{13}$ = $p\widehat{\underbar{m}}_{xy} \widehat{\underbar{2}}_{x\bar{y}} m_z$ \\
             &   &  $\bar{F}_{13}$ = $pm_z$     \\ \hline
$(1_1|~ (1\bar{1}0),p ~| 2_1)$ & $0$  &  $T_{13}$ = $p\widehat{\underbar{2}}_{xy} \widehat{\underbar{c}}_{x\bar{y}} m_z$\\ 
             &   &  $\bar{F}_{13}$ = $pm_z$ \\ 
             & $\frac{1}{2}$  & $T_{13}$ = $p m_z$  \\
             &   &  $\bar{F}_{13}$ = $p m_z$     \\ \hline
\end{tabular}}
\label{tab:tableII}
\end{table}

To calculate the symmetry of an orientational (ferroelastic) domain boundary
$(1_1| ~(1,1,0), -\frac{1}{2}~|2_1)$  we proceed as before. 
Inspecting Fig.\,\ref{fig:Domain pairs} we identify the 
symmetry operations of $ F_{13} = P2_1/m_z$, which leave $p=-\frac{1}{2}$ and $\textbf{n} = (1,1,0)$ invariant. They form the layer group $\bar{F}_{13} = \textbf{T} \{(1/000)~(m_z/000)\} \equiv pm_z$, where $\textbf{T} = n(\textbf{a}-\textbf{b})+m\textbf{c} ~ (n,m \in \mathbb{Z})$.
Those elements of $  J_{13}''$, which at $p=-\frac{1}{2}$ change $\textbf{n} \rightarrow -\textbf{n}$ are 
$\color{black} T_{13}'' = \textbf{T} \{(m_{xy}/\frac{1}{2}\frac{1}{2}0) (2_{x\bar{y}}/\frac{1}{2}\frac{1}{2}0)\}$. 
Combining both yields 
\begin{equation}
    T_{13} = {\bf T} \{(1/000)~(m_z/000)~{ \color{OliveGreen} (m_{xy}/\frac{1}{2}\frac{1}{2}0) (2_{x\bar{y}}/\frac{1}{2}\frac{1}{2}0)} \} ~.
\end{equation}
Therefore, the resulting symmetry is $T_{13} \equiv p  \widehat{\underbar{m}}_{xy} \widehat{\underbar{2}}_{x\bar{y} }  m_z$ (Table~\ref{tab:tableII}).

\subsection{Translational domain boundaries of KSCN} \label{IVB}

Unlike ferroelastic domain walls, translational anti-phase boundaries are not subject to strain compatibility relations, and can therefore generally be oriented (if we neglect other anisotropy effects) in any direction.  
Let us start with the domain boundary $(1_1/(0,1,0), 0/1_2)$. To calculate $\bar{F}_{12}$ we select those symmetry operations of $ F_{12} = Pbcm$, which at $\textbf{p} = (000)$ leave $\textbf{n} = (0,1,0)$ invariant. Inspecting Fig. \ref{fig:DP} this yields 
$\bar{F}_{12} = \textbf{T} \{(1/000)~(m_z/000)\} \equiv pm_z$, where $\textbf{T} = n\textbf{a} + m\textbf{c}~ (n,m \in \mathbb{Z}) $. 
Note that $T_{12}$ is a layer group, i.e. the translational elements $\textbf{T}$ act only parallel with the plane of the domain wall.
To determine $ T_{12}''$ we take those operations of $ J_{12}''$, which at $\textbf{p} = (000)$ change $\textbf{n} \rightarrow -\textbf{n}$. These are $ T_{12}'' = \textbf{T}\{(\bar{1}/000)~(2_z/000)\}$. Thus, 
\begin{eqnarray}
T_{12} & = &\textbf{T}\{(1/000)~(m_z/000) 
~{\color{blue}(\overline{1}/000)~(2_z/000)}\}  \nonumber \\
& \equiv & p\widehat{\underline{2}}_z / m_z ~.
\end{eqnarray}

\begin{figure}
\centering
\includegraphics[scale=0.45]{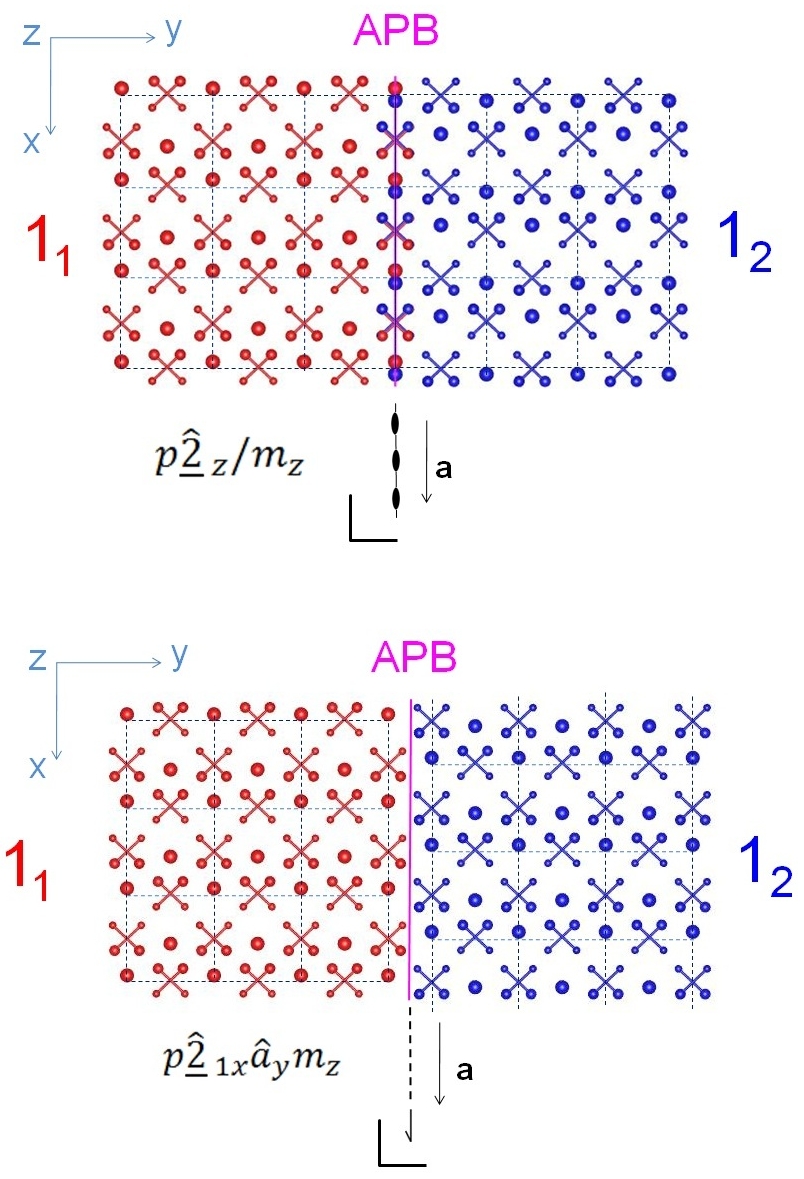}
\caption{Translational anti-phase boundaries with ${\bf n}=(0,1,0)$ orientation at positions $p=0$ and $p =\frac{1}{4} $ together with corresponding layer groups $T_{12}$.}
\label{fig:APB 010 0}
\end{figure}

For the domain boundary $(1_1|(0,1,0), \frac{1}{4})|1_2)$ (see Fig.\,\ref{fig:DP} we obtain by checking for the corresponding symmetry operations preserving the $(010)$-plane at $p\frac{1}{4}$. This yields
$T_{12}' = \textbf{T} \{(1/000)~(m_z/000)\} \equiv pm_z$ and $ T_{12}'' = \textbf{T}\{(2_x/\frac{1}{2}\frac{1}{2}0)~(m_y/\frac{1}{2}\frac{1}{2}0)\}$.
Thus, for $(1_1|(0,1,0), \frac{1}{4})|1_2)$, the $T_{12}$ layer group reads
\begin{eqnarray}
    T_{12} & = &  \textbf{T}\{(1/000)~(m_z/000)~{\color{blue}(2_x/\frac{1}{2}\frac{1}{2}0)~(m_y/\frac{1}{2}\frac{1}{2}0)} \} \nonumber \\
    & = &  p  \widehat{\underline{2}}_{1x} \widehat{\underline{a}}_y m_z ~.
\end{eqnarray}
Fig.\,\ref{fig:APB 010 0} displays the anti-phase boundaries for these two different positions $p=0$ and $p=\frac{1}{4}$ together with the corresponding symmetry groups $T_{12}$.
It is obvious, that $T_{12}$ at  $p=0$  is non-polar layer group, whereas $T_{12}$ at $p=\frac{1}{4}$ is a polar layer group, which allows for a polarization component $P_x \neq 0$ (screw axis $2_{1x}$) in the corresponding domain wall. Quite similar behaviour is also obtained for other orientations of translational anti-phase boundaries (see Table\,\ref{tab:table3}).

\begin{table} [h]
\caption{Symmetry groups of translation domain boundaries of different orientations $\textbf{n}$ at various positions $\textbf{p}$. ()$^\ast$~marks asymmetric domain boundaries, i.e. no color changing operations $\in T_{12}$ exist. (NP)=non-polar group, (P)=polar group.}
{\begin{tabular}{lccc} \hline \hline  \\
Domain boundary & Position \textbf{p} &  Layer group \\ \hline 
$(1_1 |~(100),\textbf{p} ~|1_2)$ & $(000)$  & $T_{12}$ = $p\widehat{\underbar{c}}_x c_y m_z$ \quad (NP) \\
             &   &  $\bar{F}_{12}$ = $p2_x c_y m_z$ \\ 
             & $(\frac{1}{4}00)^\ast$  & $T_{12}$ = $p2_x c_y m_z$ \quad (P)\\
             &   &  $\bar{F}_{12}$ = $p2_xc_ym_z$ \\  \hline
$(1_1 |~(010),\textbf{p}~|1_2)$ & $(000)$  & $T_{12}$ = $p \widehat{\underbar{2}}_z/ m_z$ \quad (NP)\\
             &   &  $\bar{F}_{12}$ = $pm_z$ \\ 
             & $(0\frac{1}{4}0)$  & $T_{12}$ = $p \widehat{\underbar{2}}_{1x} \widehat{\underbar{a}}_y m_z$ \quad (P)\\
             &   &  $\bar{F}_{12}$ = $pm_z$ \\ \hline
\end{tabular}}
\label{tab:table3}
\end{table}

\subsection{ Inversion domain boundaries of lacunar spinels} 

We believe that the selected example of KSCN allowed us to describe most of the  aspects of that can be encountered in symmetry analysis of domain boundaries in an arbitrary nonpolar material. To broaden the perspective with another example, let us briefly consider inversion anti-phase boundaries in lacunar spinels of the GaV$_4$S$_8$ family.

At ambient conditions, these materials have noncentrosymmetric cubic structure, $F = F\bar{4}3m ~ (T_{d}^{2})$. This structure can be understood as derived from a parent, completely filled  centrosymmetric spinel of $G_0 = Fd3m ~ (O_{h}^{7})$ symmetry. The  symmetry reduction can be described by a one-component order parameter $\eta$ which transforms as the  $A_{2u}$ pseudoscalar one-dimensional irrep \cite{Talanov2014}. In other words, the parent-child relationship correspond to an equitranslational phase transition, where the macroscopic symmetry changes from $m\bar{3}m$ to $\bar{4}3m$. There are only two domain states 1 and 2 (orientational ones), describing two possible enantiomorphic forms of the material. Symmetry reduction belongs to a nonferroelectric and nonferroelastic species, but domain states 1 and 2  differ in the sign of the piezoelectric tensor \cite{Janovec1975, Hlinka2016}.
 
 The parent symmetry group has 48 symmetry operations per primitive unit cell.
 One half of these operations are proper operations (preserving handedness), the other half is formed by the improper operations (there the determinant of the rotational part of the operation equals to -1). The pseudoscalar nature of the order parameter implies that the former set of operations forms a halving subgroup describing the symmetry of the child phase (it is an identical group for both domain states)
 and the other half represents all state-exchanging operations. Therefore, $F_{12}=F\bar{4}3m$ and  $J_{12}=F\widehat{\underbar{d}}3\widehat{\underbar{m}}$.

Let us now consider an inversion anti-phase boundary perpendicular to the tetragonal axis with ${\bf n}=(1,0,0)$ passing through the inversion center of the parent phase (Wyckoff position $c$ or $d$ with site symmetry $3m$). 
In the standard setting origin at the $\bar{4}3m$ Wyckoff position $a$,
this domain boundary would thus match the position of the diagonal plane $d$ at fractional coordinate $x=\frac{1}{8}$, so that the position vector is ${\bf p}=\frac{1}{8}{\bf n}$, $p=\frac{1}{8}$. The anti-phase domain boundary normal and domain state (handedness) are both preserved only by identity $1$, $4_x$, $2_x$ and $4_x^3$ operations, and none of these operations shifts the domain boundary $(1| (0,0,1), \frac{1}{8} |2)$, so that 
 $\bar{F}_{12}=p4_x$. 
 Simultaneous flipping of the domain boundary normal and domain state (handedness) can be accomplished by $\bar{1}$, $\bar{4}_x$, $d_x$ and $\bar{4}_x^3$. Here again, none of these operations shifts the domain boundary located at $p=\frac{1}{8}$.
 By making union of both sets, the symmetry of the $(1| (0,0,1), \frac{1}{8} |2)$ domain boundary is obtained as $T_{12}([100], (000))=p4_x/\widehat{\underbar{n}}_x$. This is a nonpolar group. On the other hand, if we assume any other position of the domain boundary, then $\bar{1}$, $\bar{4}_x^3$, $m_x$ and $\bar{4}_x^3$ are not symmetry operations any more
 and we are left with a polar symmetry layer group $T_{12} = p4_x$.

\section{Symmetry of order parameter profiles} \label{sec:V}

\subsection{Ferroelastic domain boundaries of KSCN} \label{VA}

In the following we show that a proper combination of layer groups with OP-symmetry is very useful to get a clue on the domain wall profiles of OP-components, polarization profiles, etc. even without solving Euler-Lagrange equations. 
Let us consider e.g. the example of a $(110)$-oriented ferroelastic domain wall $(1_1|~(110),-\frac{1}{2}~|2_1)$. 
The symmetry elements of $T_{13} = \textbf{T} \{1, m_z, 
\widehat{\underbar{2}}_{x \bar{y}}, \widehat{\underbar{m}}_{xy}\}$
are condensed in Table\,\ref{tab:tableIII}, together with the corresponding irreducible representations for the polarization $V(g)$ and order-parameter. 

\begin{figure}
\centering
\includegraphics[scale=0.4]{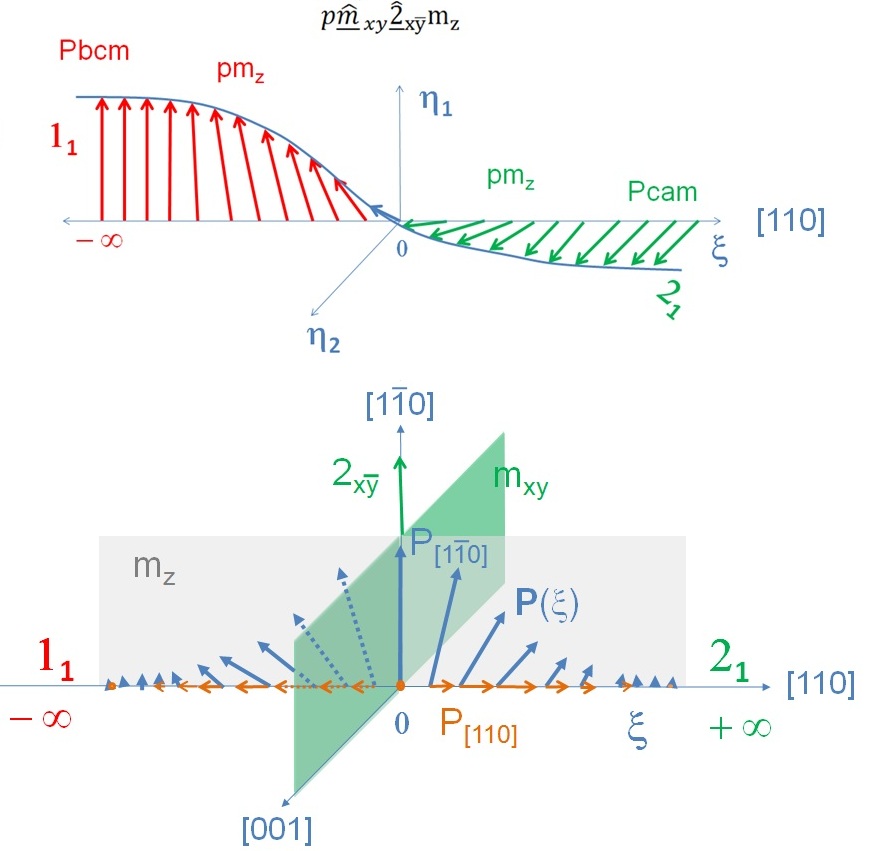}
\caption{Top: OP profile of a ferroelastic domain boundary with wall with $n=(1,1,0)$ orientation. Bottom: Polarization profile $\textbf{P}(\xi) = (P_{[1\bar{1}0]}(\xi), P_{[110]}(\xi))$, which is compatible with the symmetry $T_{13} = p \color{black}\widehat{\underbar{m}}_{xy} \widehat{\underbar{2}}_{x\bar{y}} \color{black} m_z$ of the domain boundary. Note, that $P_{[1\bar{1}0]}$ is symmetric with respect to $\xi$, whereas $P_{[110]}$ is anti-symmetric to fulfil the symmetry requirements of $T_{13}$.}
\label{fig:OP profile}
\end{figure}

\begin{table*} 
\caption{Symmetry elements of the layer group $T_{13}$ = $p\widehat{\underbar{m}}_{xy} \widehat{\underbar{2}}_{x\bar{y}} m_z$ describing symmetry of $(1_1 |(1,1,0), \frac{1}{2}|2_1)$ domain boundary and corresponding irreducible representations of the order parameter $\tau_{ij} (g)~(i,j=1,2)$ and polarization $V_{ij}(g)~(i,j=1,2,3)$.}
{\begin{tabular}{cccccc} \\
$g$ & $(1/000)$ & $(m_z/000)$ & \color{OliveGreen}$(2_{x\bar{y}}/\frac{1}{2}\frac{1}{2}0)$ &\color{OliveGreen} $(m_{xy}/\frac{1}{2}\frac{1}{2}0)$ & OP \\ \hline \hline
$\tau (g)$ & $ \begin{pmatrix} 1 & 0\\ 0 & 1 \end{pmatrix} $ & 
             $ \begin{pmatrix} 1 & 0\\ 0 & 1 \end{pmatrix} $ &
             $ \begin{pmatrix} 0 & 1\\ 1 & 0 \end{pmatrix} $ &
             $ \begin{pmatrix} 0 & 1\\ 1 & 0 \end{pmatrix} $ &
$ \left(
\begin{array}{c}
\eta_1\\
\eta_2\\
\end{array}
\right)$ \\ \hline
$V(g)$ & $ \begin{pmatrix} 1 & 0 & 0\\ 0 & 1 & 0\\ 0 & 0 & 1 \end{pmatrix} $ &
         $ \begin{pmatrix} 1 & 0 & 0\\ 0 & 1 & 0\\ 0 & 0 & -1 \end{pmatrix} $ &
         $ \begin{pmatrix} 0 & -1 & 0\\ -1 & 0 & 0\\ 0 & 0 & -1 \end{pmatrix} $ &
         $ \begin{pmatrix} 0 & -1 & 0\\ -1 & 0 & 0\\ 0 & 0 & 1 \end{pmatrix} $ &
$ \left(
\begin{array}{c}
P_x\\
P_y\\
P_z\\
\end{array}
\right)$ \\ \hline

\end{tabular}}
\label{tab:tableIII}
\end{table*}

The layer group $T_{13}$ of the domain boundary requires that the OP profile along the path $\xi \textbf{n}$  ($-\infty < \xi  < \infty$) is invariant with respect to the symmetry operations $g \in T_{13}$. Since the OP components transform under the action of a symmetry element $g$ according to $\tau (g)$ this translates to $\tau (g)(\eta_1(\xi), \eta_2(\xi))$. 
We find for all symmetry elements $f \in \bar{F}_{13}$ (black elements in Table \ref{tab:tableIII}) $\tau (f)(\eta_1(\xi), \eta_2(\xi))=(\eta_1(\xi), \eta_2(\xi))$. Symmetry operations of $T_{13}''$  change $\textbf{n} \rightarrow -\textbf{n}$, i.e. $\xi \rightarrow -\xi $, so that for them  $\tau (\eta_1(\xi), \eta_2(\xi))=(\eta_2(-\xi), \eta_1(-\xi))$. Taking it together, the symmetry of the domain boundary requires for the order parameter components the relation $\eta_1(\xi) = \eta_2(-\xi)$ and $\eta_2(\xi) = \eta_1(-\xi)$. At the domain wall centre $(\xi =0)$ these relations lead to $\eta_1(0) = \eta_2(0)$.

So, even without solving the Euler-Lagrange equations of the corresponding Landau-Ginzburg-Devonshire free energy expansion, one gets a good guess of the domain wall profile in OP-space. Fig.~\ref{fig:OP profile} shows a sketch of the OP profile in a ferroelastic domain wall of KSCN with the corresponding symmetry groups attached. Note, that only at $ \xi  = \pm \infty$ the space groups are 3 dimensional (marked by capital P in the space group symbol), whereas at the centre of the domain wall as well as in the regions near the domain wall (the "shoulders") the groups are layer groups with 2 dimensional periodicity of the OP (marked by small p in the space group symbol) which varies along $\xi$.

Additionally, this method is also helpful to get a clue on the polarization profile in the domain wall.
Any polarization vector $\textbf{P} (\xi)$ which is compatible with the domain boundary symmetry $T_{13}$ has to fulfill the condition $V(T_{13})\textbf{P}(\xi)=\textbf{P}(\xi)$. Inspecting Table \ref{tab:tableIII} this implies that for the "shoulder" region $0 < \vert \xi \vert < \infty$ two non-zero polarization components are possible in the (x,y)-plane, i.e.  $\textbf{P}(\xi) = (P_x(\xi),P_y(\xi),0)$. 
For further considerations it is instructive to split the polarization vector into a component parallel to the domain wall, i.e. $P_{[1\bar{1}0]}=(P,-P,0)$ and a component perpendicular to the domain wall, i.e. $P_{[110]}=(P,P,0)$. From Table~\ref{tab:tableIII} we find, that the symmetry elements which change $\xi$ into $-\xi$ transform $P_{[1\bar{1}0]}=(P,-P,0)$ into $(P,-P,0) = P_{[1\bar{1}0]}$, implying  $P_{[1\bar{1}0]}(-\xi) =  P_{[1\bar{1}0]}(\xi)$, i.e. a symmetric profile.    
For the component parallel to the domain wall normal $P_{[110]}=(P,P,0)$ changes to $(-P,-P,0) = -P_{[110]}$ for $\xi \rightarrow -\xi$ , implying  $P_{[110]}(-\xi) = - P_{[110]}(\xi)$, i.e. the profile is anti-symmetric. Fig.~\ref{fig:OP profile} (bottom) shows a sketch of the polarization profile of a ferroelastic domain wall which is compatible with the symmetry $T_{13}$ of the domain boundary and which clearly shows that these symmetry requirements are fulfilled.

The symmetry of the $(1_1|~(110),p~|2_1)$ ferroelastic domain boundary at the position $p=0$ is $T_{13} = p\widehat{\underbar{2}}_{xy} \widehat{\underbar{c}}_{x\bar{y}} m_z$ (see Table~\ref{tab:tableIV}). There we find that the profile of the polarization component perpendicular to \textbf{n} is symmetric, $P_{[110]} = (P,P,0) \rightarrow (P,P,0)=P_{[110]}$ for $\xi \rightarrow -\xi$, while the profile of the normal polarization component is anti-symmetric, $P_{[1\bar{1}0]} = (P,-P,0) \rightarrow (-P,P,0)=-P_{[1\bar{1}0]}$ for $\xi \rightarrow -\xi$.

\begin{table*}
\caption{Symmetry elements of the layer group $T_{13} = p\widehat{\underbar{2}}_{xy} \widehat{\underbar{c}}_{x\bar{y}} m_z$ describing symmetry of $(1_1 |(1,-1,0), 0 |2_1)$ domain boundary and corresponding irreducible representations of the order parameter $\tau_{ij} (g)~(i,j=1,2)$ and polarization $V_{ij}(g)~(i,j=1,2,3)$.}
{\begin{tabular}{cccccc} \\
$g$ & $(1/000)$ & $(m_z/000)$ & \color{OliveGreen}$(2_{xy}/00\frac{1}{2})$ &\color{OliveGreen} $(m_{x\bar{y}}/00\frac{1}{2})$ & OP \\ \hline \hline
$\tau (g)$ & $ \begin{pmatrix} 1 & 0\\ 0 & 1 \end{pmatrix} $ & 
             $ \begin{pmatrix} 1 & 0\\ 0 & 1 \end{pmatrix} $ &
             $ \begin{pmatrix} 0 & 1\\ 1 & 0 \end{pmatrix} $ &
             $ \begin{pmatrix} 0 & 1\\ 1 & 0 \end{pmatrix} $ &
$ \left(
\begin{array}{c}
\eta_1\\
\eta_2\\
\end{array}
\right)$ \\ \hline
$V(g)$ & $ \begin{pmatrix} 1 & 0 & 0\\ 0 & 1 & 0\\ 0 & 0 & 1 \end{pmatrix} $ &
         $ \begin{pmatrix} 1 & 0 & 0\\ 0 & 1 & 0\\ 0 & 0 & 1 \end{pmatrix} $ &
         $ \begin{pmatrix} 0 & 1 & 0\\ 1 & 0 & 0\\ 0 & 0 & -1 \end{pmatrix} $ &
         $ \begin{pmatrix} 0 & 1 & 0\\ 1 & 0 & 0\\ 0 & 0 & 1 \end{pmatrix} $ &
$ \left(
\begin{array}{c}
P_x\\
P_y\\
P_z\\
\end{array}
\right)$ \\ \hline
\end{tabular}}
\label{tab:tableIV}
\end{table*}

\subsection{Anti-phase boundaries of KSCN} \label{VB}

\begin{table*}[ht]
\caption{Symmetry elements of the layer groups $T_{12}$ for $ (1_1 |~(0,1,0),p| 1_2)$ domain boundaries  corresponding irreducible representations $\tau_{ij} (g)~(i,j=1,2)$ and vector representations $V_{ij}(g)~(i,j=1,2,3)$.}
{\begin{tabular}{cccccc}  \\
$p=0 $ & $(1/000)$ & $(m_z/000)$ & \color{blue}$(\bar{1}/000)$ &\color{blue} $(2_z/000)$ & OP \\ \hline \hline
$\tau (g)$ & $ \begin{pmatrix} 1 & 0\\ 0 & 1 \end{pmatrix} $ & 
             $ \begin{pmatrix} 1 & 0\\ 0 & 1 \end{pmatrix} $ &
             $ \begin{pmatrix} -1 & 0\\ 0 & -1 \end{pmatrix} $ &
             $ \begin{pmatrix} -1 & 0\\ 0 & -1 \end{pmatrix} $ &
$ \left(
\begin{array}{c}
\eta_1\\
\eta_2\\
\end{array}
\right)$ \\ \hline
$V(g)$ & $ \begin{pmatrix} 1 & 0 & 0\\ 0 & 1 & 0\\ 0 & 0 & 1 \end{pmatrix} $ &
         $ \begin{pmatrix} 1 & 0 & 0\\ 0 & 1 & 0\\ 0 & 0 & 1 \end{pmatrix} $ &
         $ \begin{pmatrix} -1 & 0 & 0\\ 0 & -1 & 0\\ 0 & 0 & -1 \end{pmatrix} $ &
         $ \begin{pmatrix} -1 & 0 & 0\\ 0 & -1 & 0\\ 0 & 0 & 1 \end{pmatrix} $ &
$ \left(
\begin{array}{c}
P_x\\
P_y\\
P_z\\
\end{array}
\right)$ \\ \hline
$p=\frac{1}{4}$ & $(1/000)$ & $(m_z/000)$ & \color{blue}$(2_x/\frac{1}{2}\frac{1}{2}0)$ &\color{blue} $(m_y/\frac{1}{2}\frac{1}{2}0)$ & OP \\ \hline \hline
$\tau (g)$ & $ \begin{pmatrix} 1 & 0\\ 0 & 1 \end{pmatrix} $ & 
             $ \begin{pmatrix} 1 & 0\\ 0 & 1 \end{pmatrix} $ &
             $ \begin{pmatrix} -1 & 0\\ 0 & 1 \end{pmatrix} $ &
             $ \begin{pmatrix} -1 & 0\\ 0 & 1 \end{pmatrix} $ &
$ \left(
\begin{array}{c}
\eta_1\\
\eta_2\\
\end{array}
\right)$ \\ \hline
$V(g)$ & $ \begin{pmatrix} 1 & 0 & 0\\ 0 & 1 & 0\\ 0 & 0 & 1 \end{pmatrix} $ &
         $ \begin{pmatrix} 1 & 0 & 0\\ 0 & 1 & 0\\ 0 & 0 & 1 \end{pmatrix} $ &
         $ \begin{pmatrix} 1 & 0 & 0\\ 0 & -1 & 0\\ 0 & 0 & -1 \end{pmatrix} $ &
         $ \begin{pmatrix} 1 & 0 & 0\\ 0 & -1 & 0\\ 0 & 0 & 1 \end{pmatrix} $ &
$ \left(
\begin{array}{c}
P_x\\
P_y\\
P_z\\
\end{array}
\right)$ \\ \hline

\end{tabular}}
\label{tab:table4}
\end{table*}

Let us consider for comparison the two translational anti-phase boundaries (Fig.~\ref{fig:APB 010 0}) with $\textbf{n}=(0,1,0)$, at $p=0$ and at $p =\frac{1}{4}$.
Table\,\ref{tab:table4} shows the symmetry elements of $T_{12}$ with corresponding irreps and vector representations, whose application leads to the following important results: The symmetry of the layer group of the translational domain boundary $(1_1 |(0,1,0),0 | 1_2)$  (Fig.~\ref{fig:APB 010 0} top) allows only for one component (e.g. $\eta_1$) of the OP to vary within the domain wall. The other component (e.g. $\eta_2$) has to be strictly zero, i.e. for $1_1 \rightarrow 1_2$ at $\textbf{p}=\textbf{0}$ the OP path along $-\infty \leq \xi \leq \infty$ must fulfil the condition $(\eta,0) \rightarrow (\eta(\xi),0) \rightarrow (-\eta,0)$. 
This results from the action of the color changing symmetry operations (Table\,\ref{tab:table4}) on the OP components, which for $\textbf{n} \rightarrow -\textbf{n}$, imply $\eta_2(-\xi) = -\eta_2(\xi) \rightarrow \eta_2(\xi=0)=0$. Such a domain wall corresponds to the straight (LAPB) path 1 in Fig.~\ref{fig:OP space}. 

For a translational anti-phase boundary at $p=\frac{1}{4}$ the situation is quite different. 
Here the layer group (second part of Table\,\ref{tab:table4}) allows for a two component OP $(\eta_1(\xi),\eta_2(\xi))$ within the corresponding domain wall. This is, because the color changing elements do not change the sign of $\eta_2(\xi)$ if $\xi \rightarrow -\xi$. Thus, for the anti-phase boundary at $p=\frac{1}{4}$  
(Fig.~\ref{fig:APB 010 0} bottom) the OP varies for $-\infty \leq \xi \leq \infty$ as $(\eta,0) \rightarrow (\eta_1(\xi),\eta_2(\xi)) \rightarrow (-\eta,0)$, via $(0,\eta_2(0))$ at the domain wall center. Such a boundary - which corresponds to the side path 2 or 2' (RAPB in Fig.~\ref{fig:OP space}).

\section{Landau-Ginzburg theory} \label{sec:VIb}

The profile of the order parameter across the domain boundary can be calculated in the framework of Landau-Ginzburg-Devonshire theory. In this approach, the  expression for the Landau-Devonshire free energy density is complemented by weakly nonlocal terms depending on spatial gradients of the order parameter components. Landau-Devonshire model for phase transition in KSCN reads \cite{Schranz1994}

\begin{eqnarray}
\label{eq:free energy}
\Phi & = & \frac{A(T-T_c)}{2}(\eta_1^2 + \eta_2^2) + \frac{B_1}{4}(\eta_1^4 + \eta_2^4) + \frac{B_2}{2}(\eta_1^2  \eta_2^2) + ... \nonumber \\
&+&  + \frac{1}{2}C_{ijkl}^0 \varepsilon_{ij} \varepsilon_{kl} + 
\Phi_{\eta,\varepsilon}(\eta_i,\varepsilon_{km})  + \Phi_{\eta,P}(\eta_i,P_j)
\end{eqnarray}

where

\begin{eqnarray}
\label{eq:coupling free energy}
\Phi(\eta_i,\varepsilon_{km})& =& a(\eta_1^2 + \eta_2^2)(\varepsilon_{11} + \varepsilon_{22}) + c(\eta_1^2 + \eta_2^2)\varepsilon_{33} \nonumber \\
&+& b (\eta_1^2 - \eta_2^2)(\varepsilon_{11} - \varepsilon_{22}) 
\end{eqnarray}

describes the lowest order coupling between strain and the order parameter and $ \Phi_{\eta,P}(\eta_i,P_j)$ contains all terms needed to describe local coupling of polarization and the order parameter.

The lowest order gradient terms in all materials include terms in the form
\begin{equation}
\label{eq:flexo}
\Phi_{\rm g} +\Phi_{\rm f}= g_{ijkl}\frac{\partial \eta_i}{\partial x_j}\frac{\partial \eta_k}{\partial x_l} + f_{ijkl}P_k \frac{\partial \varepsilon_{ij}}{\partial x_l} ~,
\end{equation}
where the first term is the usual Ginzburg term and the latter is the flexoelectric coupling \cite{Morozovska2012}. In principle, multi-component order parameters allow to construct also Lifshitz-like gradient terms $\Phi_{\rm h}$. In case of KSCN, by inspecting Table\,\ref{tab:tableI}, we can easily find that the following Lifshitz-like invariant, mixing polarization components with  gradients of the primary order parameter,
\begin{eqnarray}
\label{eq:rotopolar}
 P_x \eta_1 \frac{\partial \eta_2}{\partial y} +   P_y \eta_2 \frac{\partial \eta_1}{\partial x} - P_x \eta_2 \frac{\partial \eta_1}{\partial y} -   P_y \eta_1 \frac{\partial \eta_2}{\partial x}  \qquad 
\end{eqnarray}
 is also allowed by symmetry. Thus, in principle, this term should be included in the Landau-Ginzburg-Devonshire theory of KCSN.

\subsection{Polarization in ferroelastic domain boundaries}

In general, quantitative calculations of domain wall profiles requires not only to select the right analytic form of the Ginzburg-Landau-Devoshire functional, but also to determine all relevant material constants. Nevertheless, the simplest form of the potential involves a quartic Landau potential and $\Phi_{\rm g}$ gradient term. Assuming in addition that the order-parameter trajectory of the $(1_1|~(1,1,0), p ~|2_1)$ ferroelastic domain boundary is restricted to a linear path in the order-parameter space, one can cast the solutions of the Euler-Lagrange equation in a very simple analytic form

\begin{eqnarray}
\label{eq:tuneni}
&\eta_1(\xi)& = \frac{\eta}{2}\left( 1- \tanh \frac{\xi - \xi_0}{\delta}  \right) \nonumber \\ 
&\rm{and}&  \nonumber\\
&\eta_2(\xi)& = \frac{\eta}{2}\left(1+ \tanh \frac{\xi - \xi_0 }{\delta}  \right)~,
\end{eqnarray}
 where $\eta$ is the OP of the homogeneous domain state, $\delta$ is the thickness of the domain boundary and $\xi_0$ is the ideal center of the boundary.
 
The coupling to the polarization can be considered as a second step. One frequently invoked mechanism involves the indirect coupling through the strain.
 In order to elucidate this mechanism, it is convenient to re-express the polarization $\textbf{P}$ in rotated components $P_{[1\bar{1}0]}$, $P_{[110]}$ and $P_{[001]}$. The symmetry allowed flexoelectric couplings terms (in eq.\,\ref{eq:flexo}) can be found with the help of  (Table\,\ref{tab:tableI}):
\begin{equation}
\label{eq:flexo for 110 DW perpendicular xi}
\Phi_{\rm f} = f_{\perp} P_{[1\bar{1}0]} \frac{\partial (\varepsilon_{11}-\varepsilon_{22})}{\partial \xi}
\end{equation}
and 
\begin{equation}
\label{eq:flexo for 110 DW parallel xi}
\Phi_{\rm f} = f_{\parallel} P_{[110]} \frac{\partial (\varepsilon_{11}+\varepsilon_{22})}{\partial \xi} ~.
\end{equation}
It should be noted, that symmetry would allow also a coupling of the type $\Phi_{\rm f} \propto P_{[001]} \frac{\partial \varepsilon_{kk}}{\partial \xi}$ ($k=1,2,3$), but since all ferroelastic domain boundaries are in the $(x,y)$-plane, there is no spatial variation of $\varepsilon_{kk}$ with respect to $z$, i.e. 
$\frac{\partial \varepsilon_{kk}}{\partial z}=0$, implying $P_{[001]}=0$, in agreement with the symmetry group $T_{13}$ (contains $m_z$). 

From (\ref{eq:flexo for 110 DW perpendicular xi}) and (\ref{eq:flexo for 110 DW parallel xi}) we obtain 
\begin{equation}
\label{eq:polarization for 110 DW perpendicular xi}
P_{[1\bar{1}0]} \propto  \frac{\partial (\varepsilon_{11}-\varepsilon_{22})}{\partial \xi} 
\end{equation}
and
\begin{equation}
\label{eq:polarization for 110 DW parallel xi}
P_{[110]} \propto \frac{\partial (\varepsilon_{11}+\varepsilon_{22})}{\partial \xi} ~.
\end{equation}
Since \cite{Schranz1994}
\begin{equation}
\label{eq:spont strain from OP}
\varepsilon_{11}-\varepsilon_{22} \propto (\eta_1^2 - \eta_2^2)
\end{equation}
and 
\begin{equation}
\label{eq:spont strain from OP with plus}
\varepsilon_{11}+\varepsilon_{22} \propto (\eta_1^2 + \eta_2^2)~,
\end{equation}
we obtain
\begin{equation}
\label{eq:polarization for 110 DW along xi from OP}
P_{[1\bar{1}0]} \propto  \frac{\partial (\eta_1^2 - \eta_2^2)}{\partial \xi}
\end{equation}
and
\begin{equation}
\label{eq:polarization for 110 DW perpendicular xi}
P_{[110]}(\xi) \propto  \frac{\partial (\eta_1^2+\eta_2^2)}{\partial \xi} ~.
\end{equation}
It is easy to verify that the symmetry of the polarization profiles calculated
from the above formulas agrees with those from the earlier numerical calculations ferroelastic domain walls in KSCN\cite{Rychetsky1994,footnote}.

In principle, the flexoelectric mechanism become inactive if the strain gradients near the domain wall are considerably suppressed. Nevertheless, for the present example one can easily show, that by adding the following (symmetry invariant) coupling terms 
\begin{eqnarray}
\label{eq:coupling eta P}
\Phi_{\nabla\eta P} = h_{\parallel} P_{[110]}  \frac{\partial (\eta_1^2 + \eta_2^2)}{\partial \xi} + h_{\perp} P_{[1\bar{1}0]}  \frac{\partial (\eta_1^2 - \eta_2^2)}{\partial \xi} \qquad
\end{eqnarray}
to the free energy expansion (\ref{eq:free energy})
one obtains very similar polarization profiles, as obtained from the flexoelectric coupling. Switching off the flexoelectric coupling may lead to a decrease of the effect, depending on the values of the coupling coefficients $f_{ijkl}$ and $h$ as well as on the magnitude of the spontaneous strain, etc.  
The question which of these coupling terms is the most important one, would need a careful investigation of such coefficients, which exceeds the scope of the present paper.

\subsection{Polarization in anti-phase domain boundaries} 

Some decades ago, the phase diagram applying to anti-phase boundary states was calculated \cite{Ishibashi1976,Sonin1989,Bullbich1989} by several authors. It was shown that depending on the parameters in the Landau expansion, there exist regions in which either a unique or bistable anti-phase boundary solution is stable \citep{Sonin1989}. 
For a quantitative analysis, one has to know the parameters of a Landau expansion, and we do not have them all for KSCN. But in the present work we are only interested in the qualitative properties of domain walls, using KSCN as a toy model example. For some parameters in the Landau expansion, there exist the following exact solutions \cite{Ishibashi1976,Sonin1989} for the order parameter components $\eta_1$ and $\eta_2$ within an anti-phase domain boundary
\begin{eqnarray}
\label{eq:Rotational APB}
&\eta_1(\xi)& = -\frac{\eta}{2}\left( \tanh \frac{\xi - \xi_0 + \Delta}{\delta} + \tanh \frac{\xi- \xi_0 - \Delta}{\delta} \right) \end{eqnarray}
and
\begin{eqnarray}
\label{eq:Rotational APBbis}
&\eta_2(\xi)& = \frac{\eta}{2}\left( \tanh \frac{\xi -\xi_0 + \Delta}{\delta} - \tanh \frac{\xi - \xi_0 - \Delta}{\delta} \right)~,
\end{eqnarray}
 where $\eta$ is the OP of the homogeneous DS, $\delta$ is the thickness of the anti-phase boundary and $\Delta$ roughly determines the half width of the layer, where $\eta_2$ is about half of its maximum value. It is easily seen from Eq.(\ref{eq:Rotational APB}) that the value $\Delta=0$ corresponds to the LAPB (path 1 in Fig.~\ref{fig:OP space}) and $p \neq 0$ corresponds to a RAPB, where path 2 is obtained for e.g. $\Delta >0$ and path 2' for $\Delta < 0$. Moreover, path 2 ($\eta_2 > 0$) yields
 polarization $P>0$ and path 2' ($\eta_2 < 0$) leads to polarization $P<0$. 
 
After minimizing the free energy  
\begin{eqnarray}
\label{eq:Px(y)}
P_x(y) \propto \left(\eta_1 \frac{\partial \eta_2}{\partial y} -   \eta_2 \frac{\partial \eta_1}{\partial y} \right)   \qquad 
\end{eqnarray}
and 
\begin{eqnarray}
\label{eq:Py(x)}
P_y(x) \propto \left(\eta_1 \frac{\partial \eta_2}{\partial x} -   \eta_2 \frac{\partial \eta_1}{\partial x} \right)  \qquad 
\end{eqnarray}

it can be easily inferred from (\ref{eq:Px(y)}) and (\ref{eq:Py(x)}) that a nonzero polarization is obtained only if $\eta_2(\xi) \neq 0$ in the corresponding domain wall, i.e. for RAPB walls. For LAPB ($\eta_2(\xi)=0$) no domain wall polarization is possible. In this way, the positional dependence of the APB polarization is encoded in the OP path  (1 or 2 (2') in Fig.~\ref{fig:OP space}) connecting the two translational domain states. 

Note that in the present example, the two solutions with $\Delta \neq 0$ are symmetry-related and energetically degenerated ones because $\Delta$ plays the role of the order-parameter of a symmetry-breaking phase transition in the domain boundary at a fixed position $\xi_0$. Our conjecture is that Eqs.\,\ref{eq:Rotational APB},\ref{eq:Rotational APBbis},\ref{eq:Px(y)} and \ref{eq:Py(x)} could be also applied to the problem of two different symmetry-related positions of the domain boundary in the crystal lattice.

\begin{figure*}
\centering
\includegraphics[scale=0.75]{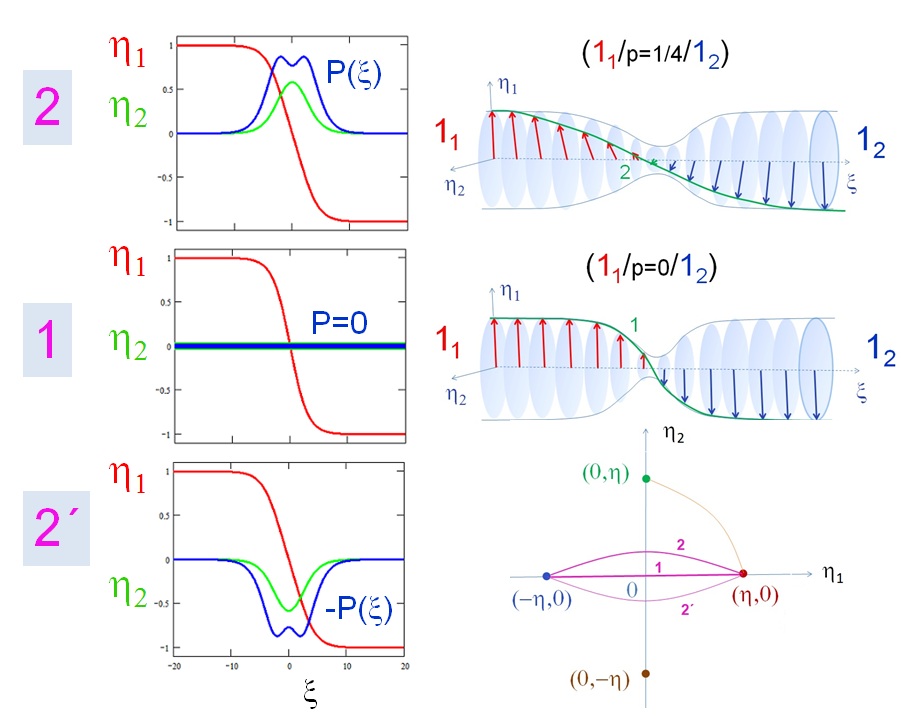}
\caption{Mid panel: Variation of the order parameter $\eta_1$ in a linear anti-phase boundary corresponding to path 1 ($\eta_2 = 0$, LAPB). Upper and lower panels show RAPB corresponding to path 2 ($\eta_2 > 0$) and path 2' ($\eta_2 < 0$). Blue lines show the corresponding anti-phase polarizations as calculated from Eqs.~\ref{eq:Rotational APB},\ref{eq:Rotational APBbis},\ref{eq:Px(y)} and \ref{eq:Py(x)}.}
\label{fig:Bloch and Ising}
\end{figure*}

\section{Polarity of Domain walls} \label{sec:VIIb}

Several examples discussed in the preceding sections are nicely illustrating the richness of the possibilities of necessary appearance of polarity in various types of domain boundaries in otherwise nonpolar crystals. In order to appreciate the overall polarity of a perfect planar domain boundary, it is sufficient to inspect the oriented crystal class (point groups $W_{ij}$) of the calculated layer group $T_{ij}$. The symmetry point group alone determines whether the symmetry-imposed polarity is present and in which direction or plane is restricted. For example, the $W_{ij}$ of mechanically compatible ferroelastic domain boundaries of KSCN indicate that their polarity is allowed and only restricted to the $m_z$ plane, except for the special positions, where the boundaries have a higher symmetry and the normal component of the polarization is vanishing. In contrast, translational anti-phase domain boundaries of KSCN can have polarization both perpendicular or parallel to the domain boundary, depending on the selected domain state pair with respect to  the otherwise equivalent orientation of the domain boundary normal, or the polarization can be completely absent, if the translational anti-phase boundary has a suitable special position. In case of (100) oriented inversion domain boundaries of lacunar spinels, the symmetry imposed polarity is always perpendicular to the boundary, except again for a special position, where the polarity vanishes completely.

Symmetry $T_{ij}$ and $W_{ij}$ does not indicate how large the domain boundary polarization could be, nor which domain wall positions will be energetically more favorable. As long as the domain boundary normal \textbf{n} is commensurate with the crystal lattice periodicity of the adjacent domain, its continuous translation along  \textbf{n}  is a process with a finite periodicity. The  energy dependence on $p$ or $\xi_0$ is then expected to have at least one minimum and one maximum on this period. The special positions of the boundary, in which the symmetry is higher than in its  general position, are potential symmetry-imposed extrema of this energy profile.
In other words, one of the special positions is likely to be the ground state domain boundary configuration, but it does not need to be so, and symmetry alone does not tell us which one it is.
Thus, the theory can typically predict the symmetry of the ground state configuration of a given domain wall uniquely only if one knows  either  its polarity, or  its groundstate position in the crystal lattice. Still, having only 2-3 candidate layer group symmetries per domain wall orientation can be helpful in any ab-initio study or transmission image analysis of domain boundary structures.

Moreover, the very existence of high-symmetry position with a forbidden polarity already implies that for this orientation of the boundary, the adjacent positions of the same boundary in polar configurations exist in symmetry related pairs, and these symmetry related locations obviously have equal energy and opposite polarization. Therefore, even if the polar configuration would be the ground state, statistical distribution of energetically preferred domain boundary positions would result in zero average polarization. 
In case of KSCN, for ferroelastic boundaries with a given domain wall normal, it implies that the statistically averaged polarization  is parallel to the boundaries. This situation is captured in our phenomenological model of Fig.\,7. Likewise, the  polarization in a set of statistically distributed parallel anti-phase boundaries  of KSCN should vanish in average, even if they have polar ground states.

The theoretical prediction of vanishing polarity in the random distribution of domain boundaries does not, however, exclude potentiality of a peculiar functional properties of materials with such anti-phase boundaries. If the ground state of the boundary is polar, and the symmetry related positions of the otherwise equivalent boundary has an opposite polarity, there is still possibility to exploit this circumstances. In a dielectric material, the external electric field should couple to the dipole moment of all boundaries and should favor localization of domain boundaries in positions 
with parallel polarity, or, in other words, the polarization can be ordered and is, in principle, switchable. These phenomena obviously assumes situations when domain boundary can be frozen at a particular position, which requires that adjacent ground state positions are separated by sufficiently high potential barriers.

This seems to be the case of the polar configurations of translational anti-phase boundaries observed \cite{Wei2014} in antiferroelectric lead zirconate (PbZrO$_3$) by electron microscopy and confirmed by ab-initio calculations. The observed shape of the in plane polarization profile of translational anti-phase boundaries in lead zirconate (red points in Figs\,.~4 and 5 of Ref.~\onlinecite{Wei2014}) is very much reminiscent of the shape (Fig.~\ref{fig:Bloch and Ising}) we obtained from the present roto-polar coupling terms Eq.~(\ref{eq:rotopolar}). 

It should be stressed, however, that this polarization switching is accompanied with nanoscale displacement of translational anti-phase boundary and in principle, one does not deal with a bistable system, but rather with an infinitely degenerate system. Thus, this type of polarization switching is a very peculiar phenomenon that deserves to be clearly distinguished from the spontaneous symmetry breaking within a fixed domain boundary, which leads to formation of degenerate domain states located at the same crystal position\cite{Sonin1989,Bullbich1989}. In this latter case, spontaneous component of the domain boundary polarization is violating the symmetry $T_{ij}$ of the boundary, similarly as for example from the Bloch-Ising  phase transitions in ferroelectric domain boundaries\cite{Stepkova2012}. 

Finally, it should be mentioned that at high temperatures, domain boundaries 
might be freely sliding within the material and in this case, the position of the domain wall within the unit cell is not a well defined quantity any more. In this case, symmetry of a sliding domain boundary can be captured by layer group that can be constructed with point group symmetries, rather than with space group crystallography. This procedure will be addressed elsewhere \cite{Janovec2019}. In either case, this complexity already indicates that  domain boundary problems deserve using several complementary approaches. We have shown that the layer group symmetry is very helpful for finding the leading order coupling terms (invariants) in a Landau-Ginzburg free energy expansion, which are needed to describe the polarization profiles of domain boundaries. In order to \textit{predict}  also the  positional dependence of polarization of domain boundaries, that is the dependence on $p$ or $\xi_0$, one would obviously need a discrete model or to include explicitly appropriate lock-in (umklapp) terms in the Landau-Ginzburg free energy expansion.

\section{Summary and conclusion}   \label{sec:Conclusion}

In a recent work Tol\'{e}dano, \textit{et al} \cite{Toledano2014} came up with a new concept to describe domain walls, which is based on order parameter symmetry only. However, in the present form of this theory one obtains domain wall structures whose symmetry is too high, and as a result the corresponding functional properties like  domain-wall polarization, etc. cannot be captured. In the present work we show that combining the theory of irreducible representations with the  layer group formalism of domain boundary symmetry yields symmetries (layer groups) of domain walls which are generally lower than those determined from the order parameter (space groups) only. We illustrate our approach on the toy example of ferroelastic domain walls and translational anti-phase boundaries in KSCN.

It should be noted that the layer group theory alone \cite{Janovec1989,Janovec2006} is already capable to describe the correct symmetries and structures of domain walls. The advantage of combining layer groups with order parameter symmetry is that in this way the concept of layer groups is readily coupled to Landau theory. It can be used to get an educated guess on the shape of order parameter profiles across a domain wall. This is probably one of the main advantages of this combined approach, since it cannot be obtained from layer groups alone, if the order parameter is e.g. a multi-dimensional quantity and not a three dimensional vector, like polarization, or simple shift of atoms, etc.
Moreover, the present approach helps a lot to identify the most important coupling terms between order parameters, order parameter gradients, polarization components, etc. in a free energy expansion, which are necessary to describe functional properties of domain walls.

Unfortunately no domain wall polarization was experimentally detected in KSCN up to now. However, we think that the present example shows that the method of combining layer groups with OP symmetry is very powerful and may help to calculate functional properties of ferroelastic domain walls and anti-phase boundaries as observed e.g. in SrTiO$_3$ \cite{Salje2013}, CaTiO$_3$ \cite{Aert2012,Yokota2017}, LaAlO$_3$ \cite{Salje2016} or PbZrO$_3$ \cite{Wei2014}, etc.\\

\noindent \textbf{Acknowledgments} It is our pleasure to thank V\'{a}clav Janovec for the deep discussions related to this work. 
One of the authors (WS) would like to thank Pierre Tol\'{e}dano for important explanatory remarks.     
The present work was supported by the Austrian Science Fund (FWF) Grant No. P28672-N36 and by the Czech Science Foundation (project no. 17-11494J).


\begin{thebibliography}{10}

\bibitem{Bain2017}
A.K. Bain, and P. Chand, \textit{Ferroelectrics: Principles and Applications}. Wiley, New York; 2017. 

\bibitem{Spaldin2012}
N. Spaldin, \textit{Magnetic Materials, Fundamentals and Applications}. Cambridge University Press; 2012.

\bibitem{Wadhawan2000}
V.K. Wadhawan, \textit{Introduction to Ferroic Materials}. Gordon and Breach Science Publishers; 2000. 

\bibitem{Tagantsev2010}
A.K. Tagantsev, L.E. Cross, and J. Fousek, \textit{Domains in Ferroic Crystals and Thin Films}. Springer, New York; 2010. 

\bibitem{Seidel2009}
J. Seidel, L.W. Martin, Q. He, Q. Zhan, Y.H. Chu, A. Rother, M.E. Hawkridge, P. Maksymovych, P. Yu, and M. Gajek \textit{et al}, 
Nature Mater. \textbf{8}, 229 (2009). 

\bibitem{Seidel2016}
J. Seidel, R.K. Vasudevan, and N. Valanoor,    
Adv. electronic mater.\textbf{2}, 1 (2016).

\bibitem{Aird1998}
A. Aird, M.C. Domeneghetti, F. Mazzi, V. Tazzoli, and E.K.H. Salje, 
J. Phys.: Condens. Matter. \textbf{10}, L569 (1998).

\bibitem{Sluka2013}
T. Sluka, A.K. Tagantsev, P. Bednyakov, and N. Setter, 
Nature Commun. \textbf{4}, 1808 (2013).

\bibitem{Aert2012}
S. Van Aert, S. Turner, R. Delville, D. Schryvers, G. Van Tendeloo, and E.K.H. Salje, 
Adv. Mater. \textbf{24}, 523 (2012). 

\bibitem{Yokota2014} 
H. Yokota, H. Usami, R. Haumont, P. Hicher, J. Kaneshiro, E.K.H. Salje, and Y. Uesu,
Phys. Rev. B \textbf{89}, 144109 (2014).

\bibitem{Yokota2017}
H. Yokota, S. Niki, R. Haumont, P. Hicher, and Y. Uesu, 
AIP Advances \textbf{7}, 085315 (2017).    

\bibitem{Salje2013}
E.K.H. Salje, O. Aktas, M.A. Carpenter, V.V. Laguta, and J.F. Scott,  
Phys. Rev. Lett. \textbf{111}, 247603 (2013). 

\bibitem{Salje2016}
E.K.H. Salje, M. Alexe, S. Kustov, M.C. Weber, J. Schiemer, G.F. Nataf, and J. Kreisel, 
Sci. Rep. \textbf{6}, 1 (2016).

\bibitem{Wei2014}
X.K. Wei, A.K. Tagantsev, A. Kvasov, K. Roleder, C.L. Jia, and N. Setter,  
Nature Commun. \textbf{5}, 3031 (2014).

\bibitem{Janovec1976}
V. Janovec, Ferroelectrics \textbf{12}, 43 (1976). 

\bibitem{Janovec1981}
V. Janovec, Ferroelectrics \textbf{35}, 105 (1981). 

\bibitem{Janovec2006}
V. Janovec, and J. P\v{r}\'{i}vratsk\'{a}, \textit{International Tables for Crystallography}, edited by A. Authier (Wiley, New York, 2006), Vol. D, Chap. 3,4, pp. 449--505.  

\bibitem{Janovec1997}
V. Janovec, and V. Kopsk\'{y},   
Ferroelectrics  \textbf{191}, 23 (1997).

\bibitem{Janovec1989}
V. Janovec, W. Schranz, H. Warhanek, and Z. Zikmund,   
Ferroelectrics \textbf{98}, 483 (1989). 

\bibitem{Kopsky2008}
V. Kopsk\'{y},  
Ferroelectrics \textbf{376}, 168 (2008).

\bibitem{Janovec2004}
V. Janovec, M. Grock\'{y}, V. Kopsk\'{y}, and Z. Kluiber,
Ferroelectrics \textbf{303}, 65 (2004).

\bibitem{Janovec2011}
V. Janovec, and D.B. Litvin,   
Phase Transitions \textbf{84}, 760 (2011).

\bibitem{Privratska1997}
J. P\v{r}\'{i}vratsk\'{a}, and V. Janovec, 
Ferroelectrics \textbf{191}, 17 (1997). 

\bibitem{Privratska2000}
J. P\v{r}\'{i}vratsk\'{a}, V. Janovec, and L. Machonsk\'{y},
Ferroelectrics  \textbf{240}, 83 (2000).

\bibitem{Ishibashi1976}
Y. Ishibashi, and V. Dvorak,  
J. Phys. Soc. Jpn. \textbf{41}, 1650 (1976). 

\bibitem{Cao1990}
W. Cao, and G.R. Barsch,  
Phys. Rev. B \textbf{41}, 4334 (1990).

\bibitem{Rychetsky1993}
I. Rychetsky, and W. Schranz,   
J. Phys.: Condensed Matter \textbf{5}, 1455 (1993).

\bibitem{Rychetsky1994}
I. Rychetsky, and W. Schranz, 
J. Phys.: Condens. Matter \textbf{6}, 11159 (1994).

\bibitem{Marton2010}
P. Marton, I. Rychetsky, and J. Hlinka,   
Phys. Rev. B \textbf{81}, 144125 (2010).

\bibitem{Wojdel2014}
J.C. Wojdel, and J. \'{I}\~{n}iguez, 
Phys. Rev. Lett. \textbf{112}, 247603 (2014). 

\bibitem{Kvasov2016}
A. Kvasov, A.K. Tagantsev, and N. Setter,
Phys. Rev. B \textbf{94}, 054102 (2016). 

\bibitem{Jiang2017}
Y.X. Jiang, Y.J. Wang, D. Chen, Y.L. Zhu, and X.L. Ma,
J. Appl. Phys. \textbf{122}, 054101 (2017). 

\bibitem{Valdez2016}
M.N. Valdez, H.T. Spanke, and N.A. Spaldin, 
Phys. Rev. B \textbf{93}, 064112 (2016).

\bibitem{Stengel2017}
A. Schiaffino and M. Stengel, Phys. Rev. Lett. \textbf{119}, 137601 (2017).  

\bibitem{Toledano1987}
C. Tol\'{e}dano and P. Tol\'{e}dano, \textit{The Landau Theory of Phase Transitions}. World Scientific Lecture Notes in Physics: Volume 3 (1987). 

\bibitem{Morozovska2012}
A.N. Morozovska, E.A. Eliseev, M.D. Glinchuk, L. Q. Chen, and V. Gopalan, 
Phys. Rev. B \textbf{85}, 094107 (2012).

\bibitem{Zubko2013}
P. Zubko, G. Catalan, and A.K. Tagantsev,
Annu. Rev. Mater. Res. \textbf{43}, 387 (2013).

\bibitem{SaljeLi2016}  
E.K.H. Salje, S. Li, M. Stengel, P. Gumbsch, and X. Ding,
Phys. Rev. B \textbf{94}, 024114 (2016).

\bibitem{Toledano2014}
P. Tol\'{e}dano, M. Guennou, and J. Kreisel, 
Phys. Rev. B \textbf{89}, 134104 (2014). 

\bibitem{Yokota2018}
H. Yokota, S. Matsumoto, E.K.H. Salje, and Y. Uesu, 
Phys. Rev. B \textbf{98}, 104105 (2018). 

\bibitem{Frenkel2017}
Y. Frenkel, N. Haham, Y. Shperber, C. Bell, Y. Xie, Z. Chen, Y. Hikita, H.Y. Hwang, E.K.H. Salje, and B. Kalisky,
Nature Mat. \textbf{16}, 1203 (2017).

\bibitem{Yamada1963}
Y. Yamada, and T. Watanabe, 
Bull. Chem. Soc. Jpn. \textbf{36}, 1032 (1963).

\bibitem{Yamamoto1987}
S. Yamamoto, M. Sakuno, and Y. Shinnaka, 
J. Phys. Soc. Jpn. \textbf{56}, 4393 (1987).

\bibitem{Zikmund1984}
Z. Zikmund, Czech. J. Phys. B \textbf{34}, 932 (1984).

\bibitem{Schranz1989}
W. Schranz, H. Warhanek, and P. Zielinsi, 
J. Phys.: Condens. Matter. \textbf{l}, 1141 (1989).

\bibitem{Schranz1994}
W. Schranz, Phase Transitions \textbf{51}, 1 (1994).

\bibitem{Kovalev1965}
O. Kovalev, \textit{Irreducible Representations of Space Groups}; New York: Gordon and Breach (1965).  

\bibitem{Ascher1977}
E. Ascher, 
J. Phys.: Solid State Phys. \textbf{10}, 1365 (1977).

\bibitem{Janovec1999}
V. Janovec, L. Richterov\'{a}  and J. P\v{r}\'{i}vratsk\'{a}, Ferroelectrics 222, \textbf{73} (1999). 

\bibitem{Bradley1972}
J.C. Bradley, and A.P. Cracknell,  1972; \textit{The Mathematical Theory of Symmetry in Solids}, Clarendon Press, Oxford.   

\bibitem{Schranz1995}
W. Schranz, \textit{Domains and interfaces near ferroic phase transitions}. Trans Tech Publications, Switzerland
(doi:10.4028/www.scientific.net/KEM.101-102.41) 
Key Eng. Mat. 1995; \textbf{101}-\textbf{102}: 41--60. 

\bibitem{Vainstein1981}
B.K. Vainstein, \textit{Modern Crystallography I}. (Springer, Berlin, 1981).
 
\bibitem{footnote}
$\eta_1^2 - \eta_2^2 = (\eta_1 + \eta_2)(\eta_1 - \eta_2) \approx (\eta_1 + \eta_2)(-1)$ and since $(\eta_1 + \eta_2)$ obeys the typical $tanh(\xi/w)$-shape, the derivative is symmetric with respect to $\xi$.

\bibitem{Sonin1989}
E.B. Sonin and A.K. Tagantsev, Ferroelectrics \textbf{98}, 291 (1989).

\bibitem{Bullbich1989}
A.A. Bullbich and Yu.M. Gufan, Ferroelectrics \textbf{98}, 277 (1989).  

\bibitem{Talanov2014}
V.M. Talanov,  and V.B. Shirokov,  Acta Cryst. \textbf{A70}, 49–63 (2014).

\bibitem{Janovec1975}
V. Janovec, V. Dvorak, and J. Petzelt,
 Czech J. Phys. B \textbf{25}, 1362 (1975).

\bibitem{Hlinka2016}
J. Hlinka, J. Privratska, P. Ondrejkovic, and V. Janovec,
Phys. Rev. Lett. \textbf{\bf 116}, 177602 (2016).

\bibitem{Stepkova2012} V. Stepkova, P. Marton, and J. Hlinka, J. Phys.: Condens. Matter \textbf{24}, 212201 (2012).

\bibitem{Janovec2019}
V. Janovec, W. Schranz and J. Hlinka, in preparation

\end{thebibliography}
\end{document}